\documentclass[floatfix,reprint, amsmath,amssymb,aps,superscriptaddress,nofootinbib]{revtex4-2}

\usepackage{graphicx}
\usepackage{dcolumn}
\usepackage{bm}
\usepackage[utf8]{inputenc}
\usepackage[export]{adjustbox}
\usepackage{wrapfig}
\usepackage{hyperref}
\usepackage{amsmath}

\usepackage{amsfonts} 
\usepackage{braket}
\usepackage{xcolor}
\usepackage{times}
\usepackage{amssymb}
\usepackage{physics}

\usepackage{xcolor}
\usepackage{amsmath}
\usepackage{gensymb}
\usepackage{textcomp}

\DeclareUnicodeCharacter{2009}{\,}
\DeclareUnicodeCharacter{03C7}{$\chi$}

\begin{document}

\title{Efficient Robust Spontaneous Parametric Down-Conversion\\via Detuning Modulated Composite Segments Designs}

\author{Muhammad Erew}
\email{erew@tauex.tau.ac.il}
\affiliation{Raymond and Beverly Sackler School of Physics and Astronomy, Tel-Aviv University, Tel-Aviv 6997801, Israel}

\author{Yuval Reches}
\email{reches@tauex.tau.ac.il}
\affiliation{Raymond and Beverly Sackler School of Physics and Astronomy, Tel-Aviv University, Tel-Aviv 6997801, Israel}

\author{Ofir Yesharim}
\affiliation{School of Electrical Engineering, Fleischman Faculty of Engineering, Tel-Aviv University, Tel-Aviv 6997801, Israel}

\author{Moshe Goldstein}
\affiliation{Raymond and Beverly Sackler School of Physics and Astronomy, Tel-Aviv University, Tel-Aviv 6997801, Israel}

\author{Ady Arie}
\affiliation{School of Electrical Engineering, Fleischman Faculty of Engineering, Tel-Aviv University, Tel-Aviv 6997801, Israel}

\author{Haim Suchowski}
\affiliation{Raymond and Beverly Sackler School of Physics and Astronomy, Tel-Aviv University, Tel-Aviv 6997801, Israel}

\date{\today}

\begin{abstract}
Spontaneous Parametric Down Conversion (SPDC) holds a pivotal role in quantum physics, facilitating the creation of entangled photon pairs, heralded single photons and squeezed light, critical resources for many applications in quantum technologies. 
However, their production is susceptible to physical variations, posing limitations on their robust utility. 
To overcome these limitations, this work introduces a method to significantly enhance the reliability of entangled photon pair generation. 
This approach involves introducing a composite design scheme to the SPDC process. 
The design is based on the development of a theoretical composite segments framework for SU(1,1), offering increased error resilience and robustness of the process. 
The practical application is experimentally demonstrated by modulating the nonlinear coefficient of a KTP crystal for degenerate 532 nm to 1064 nm conversion, resulting in an effective sevenfold improvement in stability of photon-pair generation and coincidence rate against temperature fluctuations compared to conventional quasi-phase-matching techniques. 
Furthermore, the presented concept is applicable to other physical systems that exhibit SU(1,1) dynamics. 
This methodology can create a leap forward in quantum technologies by significantly enhancing stability and error tolerance, thus paving the way for a new generation of entangled photon sources, holding promise for quantum information processing, communication, and precision measurement applications.
\end{abstract}

\maketitle

\section{Introduction}

\subsection{Background and Motivation}

\begin{figure*}
  \includegraphics[width=\linewidth]{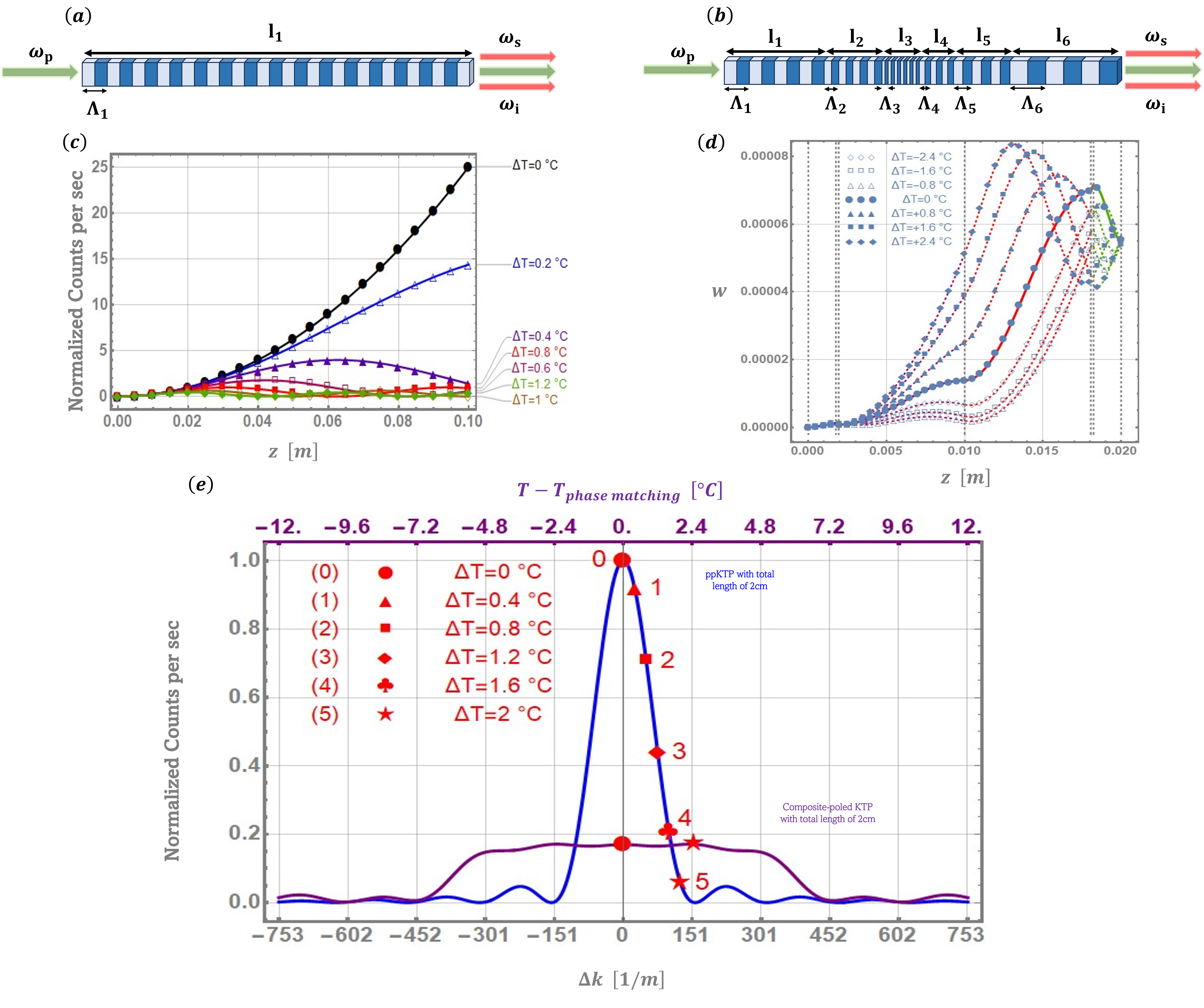}
  \caption{(a) Perfect phase matching and the periodically poled QPM design: 
  A non-linear crystal is poled periodically to compensate for the phase mismatch of the desired frequency conversion. 
  (b) DMCS crystal, in which the crystal has modulated segments with different poling periods and lengths. 
  (c) The normalized rate of generated pairs along the crystal while assuming that it lies in the Rayleigh range and the process is under plane wave approximation. 
  All dynamics in the plotted non-zero deviation temperature plots are already in the harmonic regime. 
  Yet, for short enough crystals, they do not differ from the hyperbolic perfect solution. 
  (d) We show for different temperature deviations how the normalized rate of generated pairs changes along the crystal and demonstrate that the process result is robust under these deviations. 
  (e) Blue is the generation rate of the periodically poled design and purple is the rate for DMCS. 
  As can be seen, the DMCS case is much less sensitive to $\Delta k$ (the gray horizontal axis) and, hence, to temperature (the purple horizontal axis). 
  We show perfect phase matching at $\Delta k=0$, and five representative points that are not phase-matched.}
  \label{fig:figure1}
\end{figure*}

Spontaneous Parametric Down-Conversion (SPDC) is a fundamental process in quantum 
optics, where a single photon spontaneously divides into two lower-energy 
photons \cite{Yariv1969}. This phenomenon serves as a vital resource for 
generating (heralded) single photons as well as entangled photon pairs \cite{Anwar2021} 
and squeezed light \cite{lvovsky2015}, forming the fundamental quantum resources for a myriad of groundbreaking applications in quantum cryptography \cite{Ekert1991}, quantum metrology \cite{Matthews2016}, information processing \cite{Slussarenko2019} 
and the validation of core principles within quantum mechanics \cite{Bouwmeester1997, Aspect1981}. 
Since the SPDC process should satisfy momentum conservation between the interacting photons, conventional nonlinear optical crystal designs, based on birefringent phase matching in bulk crystals or on quasi-phase-matching (QPM) in periodically poled crystals (see Figure \ref{fig:figure1}(a)), are widely used to 
obtain efficient SPDC \cite{boyd2020nonlinear,Armstrong1962,Hum2007}. 
Yet, these traditional designs are notoriously susceptible to temperature fluctuations, fabrication inconsistencies, variations in incidence angles, and alterations in pump wavelength and intensity \cite{Malara2008, Ling2008, Steinlechner2014}. 
Such sensitivities pose substantial limitations and challenges to the stability and robustness of SPDC-based photon pair generation, critically impacting the performance of quantum communication and information processing protocols in real-life scenarios.

In the past few decades, several design concepts have addressed this challenge for SPDC, including utilizing very thin bulk nonlinear crystals \cite{Okoth2019, Santiago-Cruz2021}, special metamaterial designs \cite{Guo2023,Okoth2019} and chirped aperiodically poled designs \cite{Szoke2021,Chekhova2018}. 
Chirped designs indeed offer an advantage in allowing robust and high-fidelity generation of photon pairs, yet to allow it and reduce efficiency variation across the conversion bandwidth, they require very long interaction lengths that introduce temporal walk-off and enhanced dispersion issues. 
This can lead to a temporal mismatch between the two photons, affecting the temporal characteristics of the entangled state \cite{Chekhova2018}. 
Thin bulk nonlinear crystals inherently have short propagation lengths and thus require relatively high pump intensities that, despite the short lengths, may increase parasitic nonlinearities and specifically the rate of undesired multi-pair generation processes. 
As the nonlinear coupling increases, SPDC can lead to the unwanted generation of multiple pairs of photons, which can reduce the interference visibility of the beams (a measure of their entanglement) and interfere with subsequent applications relying on their entanglement. 
Metamaterials, while promising, remain a relatively immature technology compared to more established platforms, as their designs are typically constrained by short interaction lengths, which limit generation efficiency, and by intrinsic resonances, which restrict the accessible wavelength range. 
Moreover, fabrication often suffers from low yield, and scaling beyond the chip level continues to present significant challenges.
Other custom poling methods showed the ability to shape the correlations between the signal and the idler, which is expressed by the joint spectral amplitude of the signal and idler photons \cite{Shukhin2024, Tambasco2016, Dosseva2016, Morrison2022}, but did not demonstrate robustness to variations in the system setup. 
Therefore, it may be asked whether one can design a process that can be significantly more robust than traditional solutions while maintaining pair-generation efficiency and suppressing multi-pair production.

Here, we introduce a new family of designs for SPDC that are based on a theoretical derivation of the detuning modulated composite segmentation (DMCS) for SU(1,1) (see Figure \ref{fig:figure1}(b)). 
While composite designs have found successful applications in SU(2)-based systems, such as in nuclear magnetic resonance (NMR) \cite{Shaka1985}, trapped ions \cite{Guide2003}, integrated photonics \cite{Kyoseva2019, Katzman2022,Kaplan2023}, and recently in stimulated nonlinear optical processes, their adaptation to SPDC, which exhibit SU(1,1) dynamical symmetry, remains an unexplored territory. 
More specifically, in the past five years, the advances in utilizing composite segmentation (CS) in nonlinear optics include crystal design demonstrated theoretically by Rangelov et al. \cite{Rangelov2014} and experimentally by Erlich et al. \cite{Erlich2019}. 
By implementing a Shaka-Pines-based composite segmentation \cite{Shaka1987}, a broadband and robust second harmonic generation (SHG) was obtained. 
In addition, Reches et al. showcased the effectiveness of CS in the sum frequency generation (SFG) process \cite{Reches2022}. 
In a parallel research avenue aimed at enhancing the efficiency of a wide broadband conversion, Al-Mahmoud et al. \cite{Al-Mahmoud2020} 
numerically demonstrated the suitability of the Shaka-Pines schemes for robust high-gain optical parametric amplification (OPA). 
However, developing CS schemes for SU(1,1) processes by segmentation is still scarce.

In our work, we bridge this gap by providing a systematic design framework for robust high-fidelity photon-pair generation, while maintaining high brightness, negligible multiple-pairs rate and high order generation. 
Our approach involves introducing a detuned composite design scheme to the SPDC process, based on a comprehensive theoretical composite segments framework for SU(1,1) (see Figure \ref{fig:figure1}(b)). 
By utilizing the harmonic regime of its dynamical evolution (see Figure \ref{fig:figure1}(c)), we show an increase of up to one order of magnitude in error resilience and robustness of the process (see Figure \ref{fig:figure1}(d) and (e)). 
Furthermore, we experimentally demonstrate its practical application on poled KTP crystal for degenerate 532 nm to 1064 nm conversion. 
Our measurements, which include both the photon-pair generation and the coincidence rate, show a sevenfold improvement in stability over the traditional periodically poled design. 
We also compare our results with thin nonlinear optical crystals and show that the DMCS solutions require seventeen times lower pump intensity to achieve similar peak power and temperature robustness as a periodically poled KTP crystal.

\section{Results}

\subsection{Theoretical Derivation of Detuning Modulated Composite Segmentation for SPDC}

We consider the most general three-wave mixing Hamiltonian of SPDC in a $\chi^{(2)}$ crystal in the slowly varying envelope approximation \cite{boyd2020nonlinear,yariv2007photonics}:

\begin{equation}
\begin{split}
H(z) = \hbar & \omega_s a_s^\dagger a_s + \hbar\omega_i a_i^\dagger a_i + 
       \hbar\omega_p a_p^\dagger a_p 
       \\ & + \hbar\kappa e^{-i\Delta k \cdot z} 
       a_s^\dagger a_i^\dagger a_p + \hbar\kappa e^{i\Delta k \cdot z} 
       a_s a_i a_p^\dagger
\end{split}
\end{equation}

where $\omega_s$, $\omega_i$ and $\omega_p$ are the signal, idler and pump frequencies respectively, $\omega_p = \omega_s + \omega_i$, $z$ is the position along the propagation axis, $\kappa$ is a real constant and $\Delta k$ is the effective phase mismatch parameter. 
The effective nonlinear coupling coefficient for first-order QPM is $\kappa = \frac{-2\chi^{(2)}}{\pi\hbar c}\sqrt{\frac{\omega_s\omega_i\omega_p}{n_s n_i n_p}}$, where $\chi^{(2)}$ is the effective second-order susceptibility, $n_s$, $n_i$ and $n_p$ are the refractive indices of the crystal for the signal, idler and pump, and $c$ is the speed of light in vacuum.

The Heisenberg equations of motion of the quantum operators $b_j(z) \equiv e^{i\omega_j t} a_j(z)$ ($j = s,i,p$) are the same as the classical limit of optical parametric amplification (OPA) for $A_j(z) \equiv \sqrt{\frac{n_j}{\omega_j}} E_j$ ($j = s,i,p$), where the latter are the amplitudes of the waves and $|A_j|^2$ ($j = s,i,p$) are proportional to the number of associated photons (see Appendix~\ref{app: The E.O.M. of the creation and annihilation operators}). 
The solution of these classical equations involves the Jacobi elliptic functions \cite{Baumgartner1979}. 
Under the undepleted pump approximation, where the input pump is much stronger than the signal and the idler throughout the propagation in the nonlinear crystal, the evolution of the signal and the idler amplitudes is given by:

\begin{equation} \label{eq: 2}
\frac{\partial}{\partial z} \begin{pmatrix} A_s(z) \\ A_i^*(z) \end{pmatrix} = 
\begin{pmatrix} 0 & -i\Omega e^{-i\Delta k \cdot z} \\ 
                i\Omega e^{i\Delta k \cdot z} & 0 \end{pmatrix} 
\begin{pmatrix} A_s(z) \\ A_i^*(z) \end{pmatrix}
\end{equation}

where $\Omega \equiv A_p \kappa$. These dynamical equations, which exhibit SU(1,1) 
dynamics can be written as propagators for $A_s$ and $A_i$ as follows:

\begin{equation}
\begin{pmatrix} A_s(z) \\ A_i^*(z) \end{pmatrix} = 
U(\Omega, \Delta k, z) \begin{pmatrix} A_s(0) \\ A_i^*(0) \end{pmatrix} = 
\begin{pmatrix} \alpha & \beta \\ \beta^* & \alpha^* \end{pmatrix} 
\begin{pmatrix} A_s(0) \\ A_i^*(0) \end{pmatrix}
\end{equation}

where $\alpha = e^{-i\frac{\Delta k}{2}z}\left(\cosh(gz) + i\frac{\Delta k}{2g}\sinh(gz)\right)$, $\beta = -ie^{-i\frac{\Delta k}{2}z}\frac{\Omega}{g}\sinh(gz)$ and 
$g = \sqrt{\Omega^2 - \left(\frac{\Delta k}{2}\right)^2}$. 
Note that $|\alpha|^2 - |\beta|^2 = 1$ holds by definition and is a manifestation of the Manley-Rowe relations \cite{suhara2013}, 
meaning that the number of generated signal-idler pairs is equal to the number of annihilated pump photons. 
When perfect phase-matching is reached ($\Delta k = 0$), the number of signal-idler pairs grows exponentially from vacuum while in the undepleted pump regime.

In the quantum mechanical derivation, and assuming that the pump field is an undepleted coherent state whose expected number of photons is large enough, the propagation of the operators' vector $(b_s, b_i^\dagger)^T$ in the Heisenberg picture is the same as that of the classical amplitudes $(A_s, A_i^*)^T$ in Equation 3 with the same entries of the transformation matrix. 
This implies that starting with the state $|\psi(z=0)\rangle = |n_s=0, n_i=0\rangle$, the state at position $z$ in the crystal will be $|\psi(z)\rangle = \sum_{n=0}^{\infty} \frac{1}{|\alpha(z)|} \left(\frac{e^{-i\omega_s t}\beta(z)}{e^{i\omega_i t}\alpha(z)}\right)^n |n,n\rangle$ implying a probability of $p^n(1-p)$ for $n$ pairs of idler-signal to exist at position $z$ in the crystal, where $p \equiv \frac{|\beta(z)|^2}{1+|\beta(z)|^2}$. 
As the number of expected pairs for this thermal (or geometric) distribution is $\mu \equiv \frac{p}{1-p} = |\beta(z)|^2$ and the variance is $\frac{p}{(1-p)^2} = \mu(\mu+1)$, an unavoidable generation of higher-order pairs is determined by the pump flux (see Supporting Information).
These high-order terms are usually unwanted in the use case of entangled photon pairs as they reduce the purity of the generated state. 
This traditionally sets a practical upper bound constraint on the induced pump power. 
Although the simple lowering of the pump beam power guarantees this upper bound, due to laser instabilities, which often require additional optical elements, this does not resolve the problem of sensitivity 
to variations in the system setup.

The dynamics of the SPDC process have two regimes of operation. In the case where $\left(\frac{\Delta k}{2}\right)^2 < \Omega^2$, where perfect phase matching occurs ($\Delta k = 0$), the evolution is hyperbolic. 
But, when $\left(\frac{\Delta k}{2}\right)^2 > \Omega^2$, the hyperbolic solution becomes harmonic and the exponential behavior disappears (see Figure \ref{fig:figure1}(c)). 
For short enough crystal lengths $|\beta(z)|^2$ is almost indistinguishable in both the harmonic and hyperbolic regimes. 
Therefore, in many weak-pump-power experiments, the dictated dynamics are probably harmonic, because tiny deviations in the temperature in such experiments quickly move the perfectly phase-matched process out of the hyperbolic regime.

A crucial outcome of the harmonic regime is that the expected number of signal-idler pairs along the crystal has a sinusoidal evolution. 
Contrary to the hyperbolic regime that always increases exponentially along the propagation axis until the system goes out of the undepleted regime (Figure \ref{fig:figure1}(c)), the expected number of signal-idler pairs in the harmonic regime decreases along the propagation axis, going through zero, and then increases again periodically. 
This might be used efficiently to produce single pairs of entangled signal-idler photons robustly while avoiding a high rate of multi-pair generation. 
Because of the exponentiality of the hyperbolic solution, it is hard to achieve robustness in the $\Omega$ and $\Delta k$ parameters in the hyperbolic regime, even with composite segmentation. 
On the other hand, the harmonic regime, where the evolution follows properties of a harmonic function holds promise.

in Fig.~\ref{fig:figure1}(c) we show the rate of output counts of a periodically poled (PP) crystal as a function of its length at various temperatures, where only at one of them the dependence is exponential.

\subsection{Detuning Modulated Composite Segmentation}

The composite method employs schemes with a series of segments, each having a different poling period to enable robust state transfer with a predetermined high-order pair generation rate, at the cost of the number of generated pairs for a given input power. 
This approach ensures robust SPDC and OPA processes. 
We compose segments in a way that will cancel the errors of $|\beta|^2$ up to some order - $M$ ($\beta$ of the corresponding SU(1,1) matrix is its off-diagonal element, as in Equation 3). 
Higher orders of error cancellation (larger M) result in greater robustness to variations in the physical parameters of the system. 
The error model we use here assumes that all the desired $\Delta k$ values acquire the same error, $\epsilon$, due to an error in the wave vector of the crystal. 
This error often occurs in experiments, e.g. owing to changes in pump frequency, crystal angle, or crystal temperature. 
By defining $U_j \equiv U(\Omega_j, \Delta k_j, l_j, \epsilon)$ for the $j$-th segment out of $N$, to be the matrix generated from $\Omega_j$, $\Delta k_j$, $l_j$ as in Equation 3, we obtain a composite segmentation in SU(1,1):

\begin{equation}
\begin{split}
U^{(N)}(\epsilon) & = U^{(N)}(\{\Omega_j\}_{j=1}^N, \{\Delta k_j\}_{j=1}^N, \{l_j\}_{j=1}^N, \epsilon) \\ & = U_N \ldots U_2 U_1
\end{split}
\end{equation}

To achieve robustness, we find N and $\{\Omega_j\}_{j=1}^N, \{\Delta k_j\}_{j=1}^N, \{l_j\}_{j=1}^N$ such that $\frac{\partial |\beta(\epsilon)|^2}{\partial \epsilon} = 0$, 
$\frac{\partial^2 |\beta(\epsilon)|^2}{\partial \epsilon^2} = 0$, 
$\frac{\partial^3 |\beta(\epsilon)|^2}{\partial \epsilon^3} = 0$, etc. 
We verify that the value of $|\beta(0)|^2$ is proportional to the rate of generated photons, is at least an order of magnitude less than that of a perfectly phase-matched crystal with a similar length.

Each design can be scaled by multiplying the lengths by a positive scaling parameter $r$ and dividing $\{\Delta k_j\}_{j=1}^N$ by $r$ and the incident pump power by $r^2$. 
For example, to impose sufficient robustness to temperature variations, we demand a robust design around temperature $T_p$ with a deviation of up to $T_m$, that will follow the constraint

\begin{equation}
\frac{1}{2T_m} \int_{T_p-T_m}^{T_p+T_m} \left(\frac{\mu(T)}{\mu(T_p)} - 1\right)^2 dT < 0.01
\end{equation}

and we demand that $\mu(T_p)$ is no more than an order of magnitude smaller than $\mu$ of a perfectly quasi-phase-matched crystal with the same total length to maintain acceptable conversion efficiency. 
We omit designs with $\{\Delta k_j\}_{j=1}^N$ values that the manufacturing process cannot produce due to limitations on the 
poling domain size. 
See Figure \ref{fig:figure1}(b) for an illustrative example of the implementation of this segmentation method in poled nonlinear crystals. 
We note that the current imposed metric, constraints and scaling can be changed to meet the demands of different systems and setups. 
While we presented the implementation of the DMCS scheme for the design of poled crystals, which are commonly used for pair generation via SPDC, it is also applicable to other systems that exhibit SU(1,1) dynamics such as traveling wave amplifiers for superconducting qubits \cite{Ranadive2022}. 
In addition, it can also be implemented via different segmentation methods, including temperature segmentation \cite{Rozenberg2019} or by introducing angles between different crystal segments.

\subsection{The DMCS-Crystal Design and Simulations}

Figure \ref{fig:figure1}(d) presents the change in the signal/idler count rate along the propagation in the crystal for different temperatures. 
It showcases the advantage of our design: while the number of generated pairs along the design differs and diverges when the temperature deviates by up to 5°C, they converge at the end of the crystal, making the output robust under such errors. 
in Fig.~\ref{fig:figure1}(e), we show the case of perfect (quasi) phase matched KTP crystal with length of 20 mm at $T$ = 37°C, illuminated by 532 nm pump. 
As can be seen, a deviation of only 0.445°C (or equivalently 10.64 nm in the signal's wavelength, or only 0.227° in the measurement angle of the output signal/idler) will lower the rate of generated photons to 90\% of the rate achieved by the error-free setup. 
This reduction results from moving to the harmonic regime where $\Omega^2 < \left(\frac{\Delta k}{2}\right)^2$. 
For more details on the sensitivity of $\Delta k$ see Appendix~\ref{app: Sensitivity of SPDC}.

\begin{figure*}
  \includegraphics[width=\linewidth]{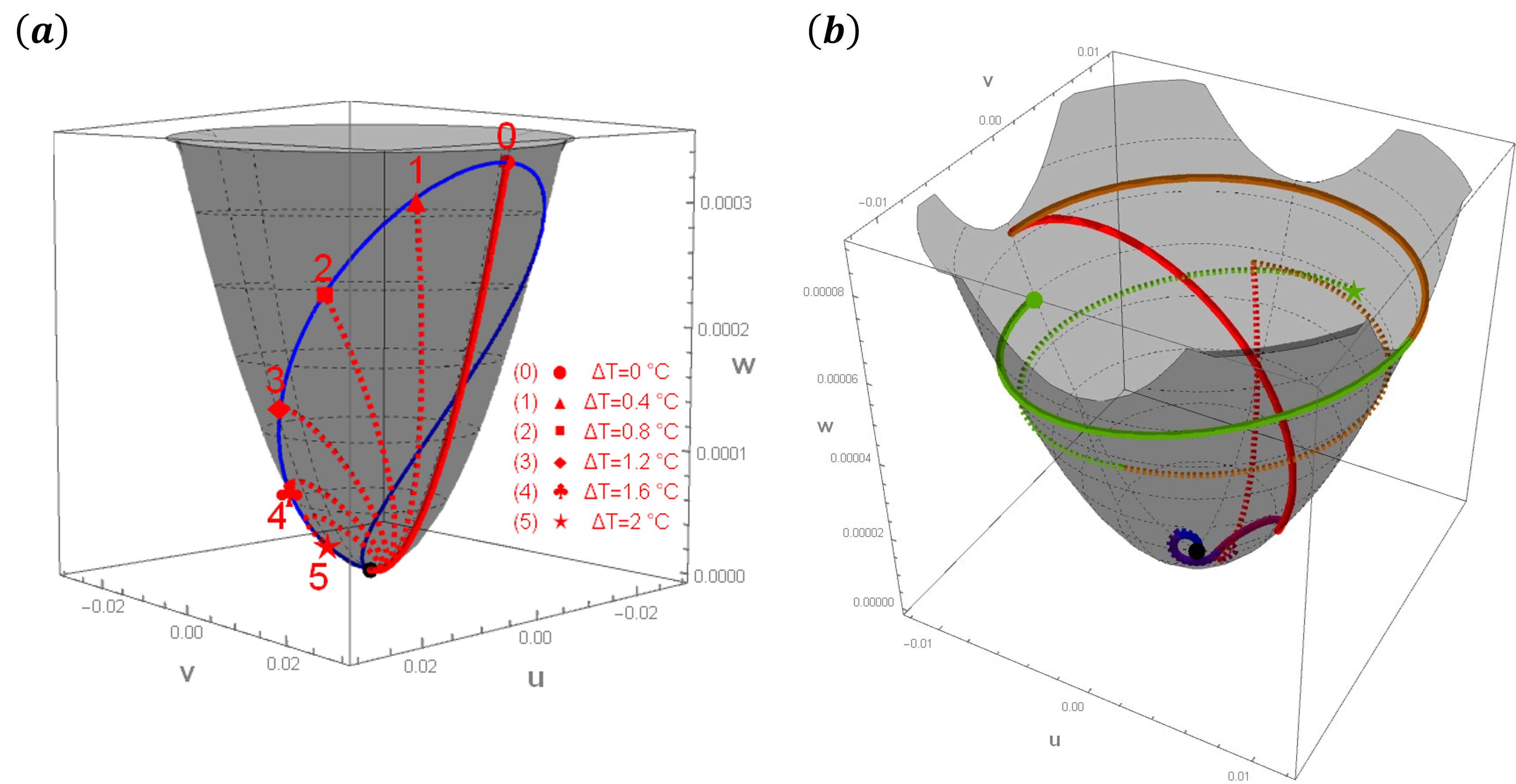}
  \caption{Geometrical representation of SPDC dynamics: All the possible SPDC states are mapped to a surface of a hyperboloid, and the pair generation rate is proportional to $w$. 
  (a) The geometrical representation of the SPDC process under the undepleted pump approximation for PP design. 
  All six path trajectories from 1(c) are plotted. The continuous red line is the perfect phase matching, and all of the dotted lines are at the five deviated temperatures. The blue continuous line describes $w$, at the end of the process as $\Delta T$ changes from -2.4°C to +2.4°C. 
  (b) The quantum $uvw$-vector trajectory on the hyperboloid representation of the SU(1,1) dynamics of our 2 cm DMCS crystal at the designated temperature (the continuous lines) and at 1°C deviation from it (the dotted lines). 
  As one can see, while the $w$ value at the end of the crystal is 6.7 times lower than the PP crystal (for the same pump intensity), it is much more robust to errors: $w(\Delta T = 0\text{°C}) = 5.545 \cdot 10^{-5}$, $w(\Delta T = 1\text{°C}) = 5.398 \cdot 10^{-5}$ (not in the figure) and $w(\Delta T = 2.4\text{°C}) = 5.579 \cdot 10^{-5}$.}
  \label{fig:figure2}
\end{figure*}

The evolution or propagation of the SPDC process (and any other SU(1,1) process) can be represented as a vector $(u, v, w)$ on the surface of a hyperboloid, in a similar manner to the Bloch sphere representation for SU(2) process \cite{Suchowski2008}. 
One can define an operator vector $(\hat{u}, \hat{v}, \hat{w})$ whose expectation values $u \equiv \langle\hat{u}\rangle$, $v \equiv \langle\hat{v}\rangle$, $w \equiv \langle\hat{w}\rangle - 1$ trace the hyperboloid $u^2 + v^2 - (w + 1)^2 = -1$ (see Appendix~\ref{app: Quantum $(u,v,w)$ Vector and Bloch Hyperboloid for SPDC} for more details), where $w$ is proportional to the pair generation rate. This is a linearized version of what has been done in \cite{Luther2000}. in Fig.~\ref{fig:figure2} we show the geometrical representation of the quantum SU(1,1) dynamics associated with the undepleted SPDC process. 
in Fig.~\ref{fig:figure2}(a) we show the trajectories of the evolution along the 2 cm PP QPM crystal, whereas in Fig.~\ref{fig:figure2}(b) we show the trajectories of the DMCS evolution. 
These are associated with the trajectories shown in Fig.~\ref{fig:figure1}(c) and Figure \ref{fig:figure1}(d), respectively. In the continuous red line, we show the trajectory of the state evolution in the perfectly phase-matched temperature, whereas in the dotted red lines we illustrate the state evolution trajectories for five other representative temperatures. 
The blue continuous line describes $w$ at the end of the process as $\Delta T$ changes from -2.4°C to +2.4°C. 
As can be seen, the process is sensitive to small deviations in the temperature, and loses about 40\% of its pair generation rate for 0.2°C deviation. 
in Fig.~\ref{fig:figure2}(b) we present the robustness of the process in our 2 cm DMCS design by depicting the quantum $(\hat{u}, \hat{v}, \hat{w})$ vector trajectory on the hyperboloid. 
The continuous lines depict the trajectory of the state evolution in the designated work temperature (i.e. $T_0 = 37$°C), whereas the dotted lines illustrate the state evolution trajectories for a 1°C deviation from it. 
The different line colors represent the different poling segments. 
The sensitivity of the process to such error or deviation is seen via the $w$ value at the output of the crystal, and one can see that the DMCS crystal did not change its $w$ value significantly: while the $w$ value at the end of the crystal is 6.7 times lower than the PP crystal (for the same pump intensity), it is much more robust to errors; 
$w(\Delta T = 0\text{°C}) = 5.545 \cdot 10^{-5}$, $w(\Delta T = 1\text{°C}) = 5.398 \cdot 10^{-5}$ (not in the figure) and 
$w(\Delta T = 2.4\text{°C}) = 5.579 \cdot 10^{-5}$. 
Thus, it is more robust than the PP QPM crystal, which changed its $w$ value much more drastically in reaction to a similar deviation (see Figure \ref{fig:figure1}(e)).

We compare our DMCS scheme to the industry benchmark, the type-0 phase-matched periodically poled design. 
The comparison utilizes a "robustness width" measurement, which is the temperature range in which the photon pairs generation rate decreases to 90\% of its maximum. 
in Fig.~\ref{fig:figure1}(e), we demonstrate the robustness of the DMCS design compared to a PP crystal. 
We show the generation rate of entangled pairs at the output of the PP crystal (in blue) and the DMCS crystal (in purple) as a function of the deviation from the perfect temperature for the process (the purple horizontal axis). 
We also present the value of $\Delta k$ (in the gray horizontal axis).

Following the design procedure, we find several designs for robust type-0 SPDC of 60 mW, 532 nm incident pump with beam radius of 40 $\mu$m shined through a KTP crystal at work temperature of $T_0 = 37$°C (the designs can be found in the Appendix~\ref{app: Selected Results}). 
The crystals were designed for degenerate SPDC, where the signal and idler have a wavelength of 1064 nm and both propagate at an angle of $\theta = 0$ relative to the optical z-axis. 
Here we focus on a design with 6 anti-symmetric segments: $\Delta k_6 = -\Delta k_1$, $\Delta k_5 = -\Delta k_2$, $\Delta k_4 = -\Delta k_3$, and the length of every segment with a phase mismatch of $\Delta k_i$ is $L_i = \frac{\pi}{2\sqrt{\left(\frac{\Delta k_i}{2}\right)^2 - \Omega^2}}$. 
We choose these simple constraints due to the analytical simplicity of finding robust designs when they are applied. 
Further details on the analytical formulation of the effect of manufacturing errors and machinery limitations on the design and poling, including the expansion of this scheme to other even number of segments, can be found in Appendix~\ref{app: Detuning Modulated Composite Segments for Robust SPDC and OPA Processes}.

\subsection{Experimental Results}

\begin{figure*}
  \includegraphics[width=\linewidth]{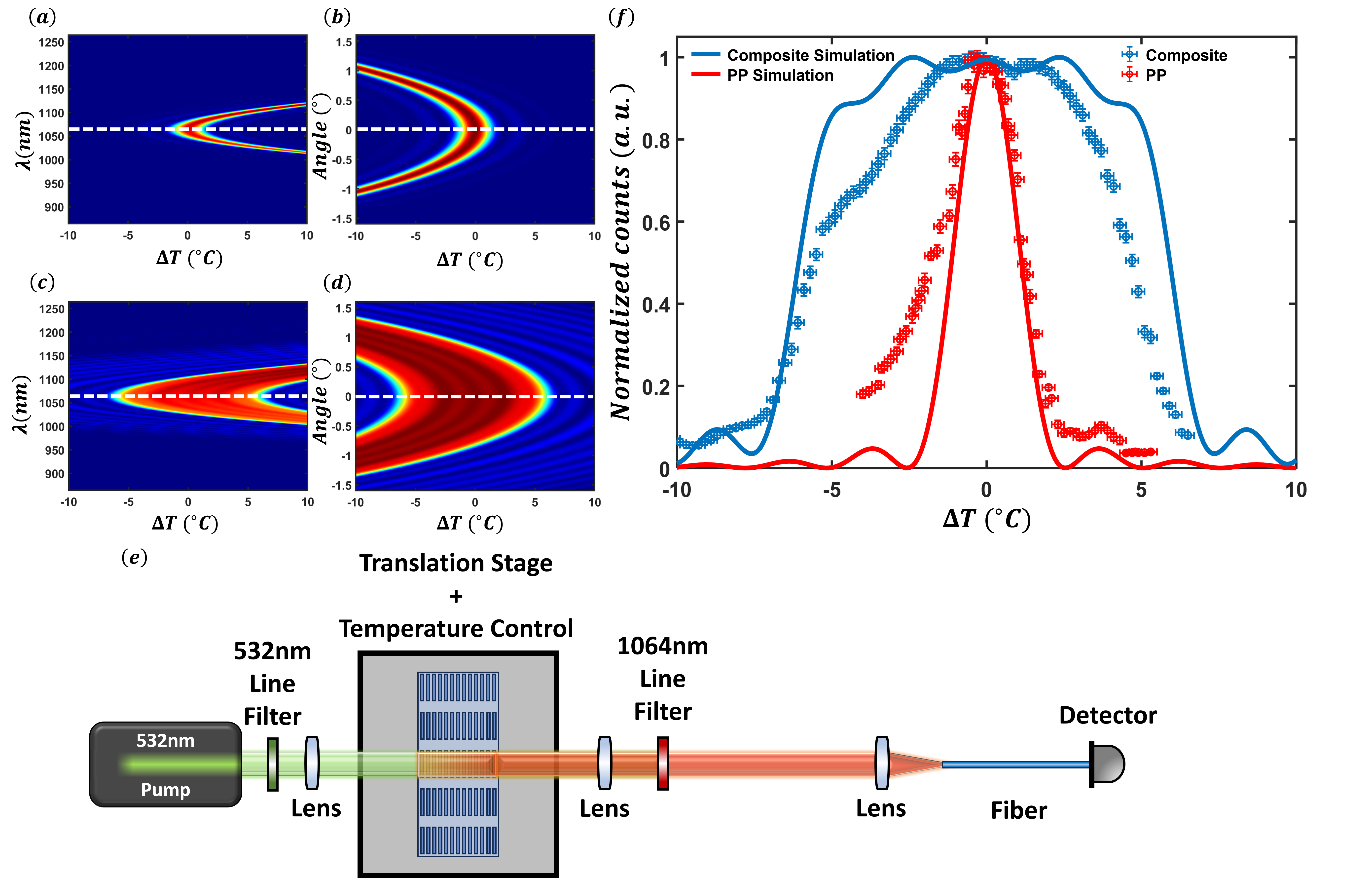}
  \caption{Comparison of numerical and experimental results for a PP crystal and a DMCS crystal. 
  (a)-(d) normalized count-rate vs. temperature deviation from the desired work temperature and the signal wavelength ((a) and (c)) or signal angle ((b) and (d)) for the PP ((a)-(b)) and DMCS ((c)-(d)) crystals, the white dashed line matches the conditions in the experiment. 
  (e) the experimental setup - a 532.25 nm CW pump laser passes through a nonlinear crystal that creates a type-0 colinear SPDC process. 
  The crystal rests on a kinetic stage that allows switching between different poling designs, and its temperature is controlled with a precision of 0.1°C. 
  After the light passes through the crystal, the remaining pump is filtered out. 
  The generated SPDC photons are focused into a fiber that collects them into a superconducting nanowire single photon detector. 
  (f) experimental and numerical results - the normalized count rate of PP (red) and DMCS (blue) crystals relative to their maximal rates versus the temperature deviation from the desired work temperature. 
  The markers are the experimental results and the continuous line marks the numerical simulations predictions.}
  \label{fig:figure3}
\end{figure*}

We compare the robustness of the PP and DMCS crystals to different physical parameters in Fig.~\ref{fig:figure3} (a)-(d). 
According to our numerical simulation results, the DMCS design should have a robustness width greater by eight and a half times than the width of the periodically poled crystal, a fourfold increase in robustness to incidence-angle variations, and an almost threefold increase in robustness to signal wavelength variation. 
We designed an experiment to verify the results of our simulations. 
A schematic of the experimental system is shown in Fig.~\ref{fig:figure3}(e). 
In the experiment, a CW pump laser with a wavelength of 532.25 nm, power of 80 mW, and beam waist of 40 $\mu$m pumped a 2 cm long, patterned KTP crystal. 
The crystal, manufactured by Raicol Crystals LTD, contains the DMCS design along with a periodically poled design for type-0 collinear SPDC. 
After passing the crystal, the beam - containing the signal and idler - passed through a 3 nm bandpass filter that removed the pump together with 
signal and idler photons far from the degeneracy wavelength, and was then coupled into a single-mode fiber and through it to a superconducting nanowire single-photon detector (SNSPD) with detection efficiency above 75\%. 
in Fig.~\ref{fig:figure3}(f) we present the experimental SPDC outcome as a function of the crystal temperature, which was controlled with a thermoelectric cooler (TEC) and a thermometer to a precision of 0.1°C. 
The composite design was found to have a width of 5°C at 90\% of its peak count rate, more than a sevenfold increase compared to the 0.7°C width of the periodically poled crystal. 
The peak count rate of the conversion process was 138.5 kcps ($10^3$ counts per second) for the DMCS design, near a twofold decrease compared to 306 kcps at the peak of the periodically poled design.
It is important to emphasize that the relevant measurement is the \textit{relative} count-rate ratio, which reflects the intrinsic characteristics of the different designs, rather than the absolute count-rates that are more susceptible to system-level optical losses.
We note that the absolute count rates are reduced due to optical losses in our setup; however, since both DMCS and PP crystals were measured under identical conditions, the relative comparison remains fully valid.
Considering manufacturing errors, measurement noise, and system imperfections, the experimental results fit well with the prediction 
of the numerical simulations. 
In addition to that, we confirmed both numerically and experimentally that the count rate is linear in the pump intensity (further details can be found in Appendices \ref{app: Errors in The Pump's Intensity} and~\ref{app: Different Pump Powers and Focusing Conditions} and in Figs.~\ref{fig: S7 errors in kappa} and~\ref{fig: S8 nlinearity and BW equation}).
We numerically compared the performance of our 2 cm DMCS design to a thin 2 mm PP KTP crystal with similar robustness width and found that the thin crystal needed seventeen times higher pump intensity to achieve a similar count rate under similar focusing conditions. 
Even after accounting for the shorter Rayleigh range of the thin crystal, which permits tighter focusing, it still required approximately twice the pump intensity (further details are also found in Appendix~\ref{app: Different Pump Powers and Focusing Conditions} and in Fig.~\ref{fig: S8 nlinearity and BW equation}).
We also note that achieving such tight focusing in practice introduces challenges such as increased alignment sensitivity, higher angular divergence, and reduced tolerance to incidence-angle drifts.

\begin{figure*}
  \includegraphics[width=\linewidth]{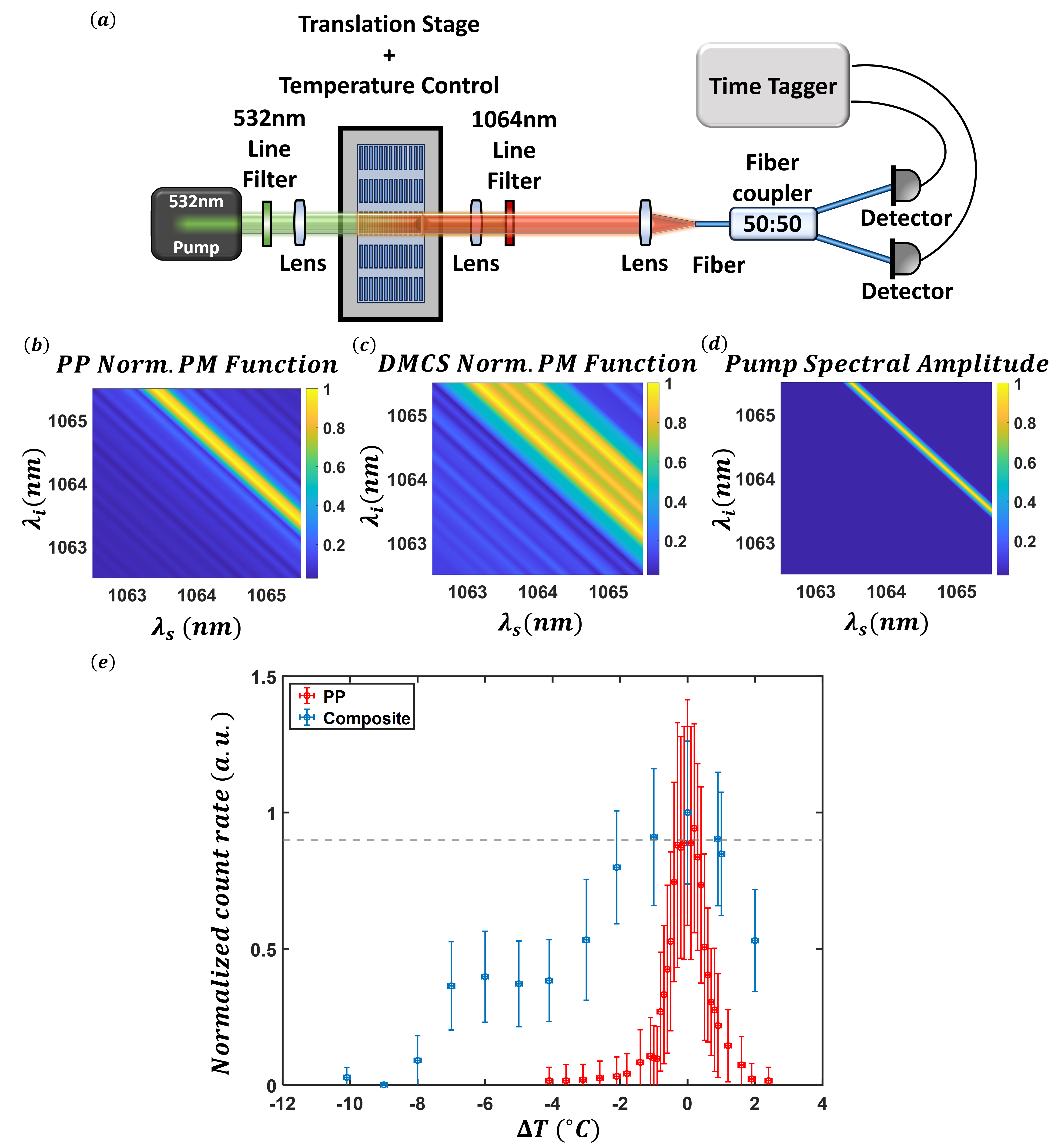}
  \caption{Comparison of the normalized coincidence rate of the PP and DMCS crystals. 
  (a) the experiment setup - we add a 50:50 fiber coupler and a second detector that allows counting coincidence between the outputs, using a time-tagger. 
  (b)-(c) normalized absolute value of the phase matching function of periodically poled and DMCS crystals inside a 3 nm wavelength window matching the line filter used in the system. 
  (d) The absolute value of the normalized pump spectral amplitude, the JSI of the crystals can be calculated by multiplying the phase matching function with the pump spectral amplitude. 
  The wider phase matching function allows the DMCS crystal to maintain a stable sum over the JSI when the temperature changes and the phase matching function shifts along the diagonal. 
  (e) experimental results - the normalized coincidence count rate of the DMCS (blue) and PP (red) crystals. 
  The DMCS crystal displays a 2°C robustness width which is almost a seven-fold improvement compared to 0.296°C of the PP crystal.}
  \label{fig:figure4}
\end{figure*}

We also measured and compared the coincidence rate of the generated photon pairs from the PP and DMCS crystals. 
For that, we added a 50:50 fiber coupler to the output fiber, as depicted in Fig.~\ref{fig:figure4}(a), and counted, using a time tagger, the coincidence rate between two single-photon detectors at 
both outputs. 
In Fig.~\ref{fig:figure4}(b) and (c) we present the normalized absolute value of the phase matching function of the PP and DMCS crystals. 
Figure \ref{fig:figure4}(d) shows the normalized absolute value of the pump spectral amplitude. 
The joint spectral intensity (JSI) \cite{Shukhin2024, Tambasco2016, Dosseva2016, Morrison2022} can be calculated by multiplying the phase matching function with the pump spectral amplitude. 
This is the absolute value squared of the joint spectral amplitude. 
The 'three stripes' shape of the phase matching function allows the DMCS crystal to maintain robustness to temperature variations that shift it diagonally, perpendicularly to the stripes. 
The JSI of both crystals at different temperatures is presented in Appendix~\ref{app: Theoretical Joint Spectral Intensity Comparison}. 
The normalized coincidence rate for both crystals is displayed in Fig.~\ref{fig:figure4}(e). 
The PP had a maximal coincidence rate of 6.12 cps and a robustness width of 0.296°C, while the DMCS crystal had 14.62 cps and 2°C. 
Here as well we find almost seven-fold improvement in the robustness of the DMCS design compared to that of the PP crystal. 
The higher coincidence rate of the DMCS crystal may stem from better alignment and coupling of the pairs into the fiber. 
Since we are measuring the robustness width relative to the maximal count rate of each crystal, the effect on it should be minimal.

\subsection{Discussion}

A central consideration in designing quantum photon sources is the balance between generation efficiency and robustness.
We note that our DMCS design explicitly prioritizes robustness to physical parameter variations while maintaining efficiency. 
This is a trade-off that is often more valuable in quantum information applications where stability and predictability are critical. 
Unlike approaches that rely on active stabilization, such as temperature control or feedback loops, which add power consumption and mechanical complexity, our method achieves robustness through passive structural elements. 
Importantly, since the SPDC process scales linearly with pump power under the undepleted pump approximation, higher generation rates can be readily obtained by increasing pump power without compromising robustness. 
This property allows our approach to achieve flexible performance optimization while offering a more favorable efficiency–robustness trade-off compared to short PP designs.\\

We also compare our DMCS design with alternative robust strategies, particularly chirped or aperiodically poled quasi-phase-matching structures (See detailed comparison Appendix~\ref{app: Comparison With Chirped Design}). 
In short, chirped designs achieve robustness via continuous change in phase-matching but typically require longer crystals, which introduce temporal walk-off, dispersion, and non-uniform spectral phase characteristics that can degrade entanglement quality. 
In contrast, our DMCS approach provides high robustness in shorter crystals, generates photon pairs more uniformly along the propagation length, and maintains efficiency even at low pump powers. 
Quantitative simulations (presented in Fig.~\ref{fig: S10 chirped_comparison} in Appendix~\ref{app: Selected Results}) show that while chirped designs can achieve broad spectral acceptance, they often exhibit significant count-rate variations and reduced efficiency at lower pump powers.
We also analyze the phase stability as a function of temperature for both design schemes. Simulation results, presented in Fig.~\ref{fig: S11 1DcomparisonWithChirp}(b) in Appendix~\ref{app: Errors in The Pump's Intensity}, show an unstable phase dependence on temperature in the chirped design. By comparison, DMCS designs deliver more uniform performance across a wider temperature range with smaller efficiency fluctuations and stable phase that can be readily compensated in conventional methods, making them particularly well-suited for compact, low-power, and stability critical applications. \\

We also note that an intriguing perspective on our work relates to the mathematical framework of pseudo-Hermitian Hamiltonians. The dynamics of SPDC under the undepleted pump approximation, as presented in Equation \ref{eq: 2}, can indeed be viewed as a special case of a pseudo-Hermitian system. 
This connection opens promising avenues for future research. As demonstrated in a paper by Torosov and Vitanov \cite{Torosov2017}, solutions developed for Hermitian Hamiltonians can be systematically translated to pseudo-Hermitian systems through appropriate transformations. 
This mathematical bridge could potentially allow us to adapt the extensive library of composite pulses and other quantum control methods originally developed for SU(2) systems to our SU(1,1) framework. 
Future work could explore this connection to develop even more sophisticated DMCS designs for SPDC, as well as extend our approach to other nonlinear optical processes, for example, difference frequency generation. 
This perspective further reinforces the broader impact of our work beyond the immediate application to robust entangled photon generation. \\

\section{Conclusion}

In this paper, we have presented a design technique for nonlinear crystals that produce entangled photons via SPDC with high robustness and efficiency. Our design solutions are based on detuning modulated composite segmentation, a new theoretical framework that applies the concept of composite pulses to the SU(1,1) dynamics. 
While we have implemented this method in SPDC, it is universal for the SU(1,1) framework. 
For example, it might be applicable for traveling wave amplifiers for superconducting qubits \cite{Ranadive2022}. 
In addition, the method is very general for SPDC and can be implemented not only via detuned segmentation of QPM but also with temperature segmentation \cite{Rozenberg2019} or by introducing angles between different crystal segments. 
We have shown in simulations that our designs can achieve a fourfold increase in the robustness to angle variations and more than an eightfold increase in the robustness to temperature variations compared to the conventional PP scheme. 
We have also demonstrated our method experimentally using a patterned KTP crystal for degenerate 532 nm to 1064 nm type-0 SPDC and found a sevenfold increase in robustness to temperature variations.
The experimental results fit well with the prediction of the numerical simulations, with the relative performance being independent of the specific system losses.
Our results indicate that our design method can significantly enhance the stability and error tolerance of entangled photon generation, which is crucial for various applications in quantum technologies. 

Additionally, we compared our DMCS design to chirped designs and showed numerically that under crystal length and pump power constraints, the DMCS design can achieve a ~45\% higher robustness width at 90\% efficiency. 
Also, the DMCS design is superior for applications that require phase uniformity because its spectral and temperature-dependent phase can be easily compensated for using common techniques.

An intriguing perspective is the connection of our framework to pseudo-Hermitian Hamiltonians, which may enable adapting composite pulse techniques from SU(2) quantum control to SU(1,1) systems. 
This opens promising avenues for extending DMCS to alternative sophisticated designs and to other nonlinear processes, further broadening the impact of our work.
We believe our approach will inspire further exploration of robust quantum light sources and open new avenues for integrating stability-driven designs into the broader landscape of quantum optics and quantum information science.

\section*{Conflict of Interest}

Prof.~Suchowski, Mr.~Erew, and Mr.~Reches hold partial rights to profit from a patent application on the design method described in this paper.  
Prof.~Goldstein, Prof.~Arie and Mr.~Yesharim declare no potential conflict of interest.

\section*{Acknowledgements}

Funding: ISF 3117/23, ISF 3427/21, Quant. Comm. Consortium of the Israel Innovation Authority.

\medskip

\bibliographystyle{MSP}
\bibliography{references}

\appendix

\renewcommand{\thefigure}{A.\arabic{figure}}
\setcounter{figure}{0}

\section{Theoretical Complementaries for SPDC}
\label{app: Theoretical Complementaries for SPDC}

\subsection{The E.O.M. of the creation and annihilation operators}
\label{app: The E.O.M. of the creation and annihilation operators}

The Heisenberg equations of motion of the signal, idler and pump operators are
\begin{subequations}
\begin{gather}
\partial_z a_s\left(z\right)=-i\omega_s a_s\left(z\right)-i\kappa e^{-i\Delta k z} a^\dagger_i\left(z\right) a_p\left(z\right),\\
\partial_z a_i\left(z\right)=-i\omega_i a_i\left(z\right)-i\kappa e^{-i\Delta k z} a^\dagger_s\left(z\right) a_p\left(z\right),\\
\partial_z a_p\left(z\right)=-i\omega_p a_p\left(z\right)-i\kappa e^{i\Delta k z} a_s\left(z\right) a_i\left(z\right),
\end{gather}
\end{subequations}
and defining $a_j\left(z\right)\equiv e^{-i\omega_j t} b_j \left(z\right)$ (for $j=s,i,p$) the equations are simpler:
\begin{subequations}
\begin{gather}
\partial_z b_s\left(z\right)=-i\kappa e^{-i\Delta k z} b^\dagger_i\left(z\right) b_p\left(z\right),\\
\partial_z b_i\left(z\right)=-i\kappa e^{-i\Delta k z} b^\dagger_s\left(z\right) b_p\left(z\right),\\
\partial_z b_p\left(z\right)=-i\kappa e^{i\Delta k z} b_s\left(z\right) b_i\left(z\right).
\end{gather}
\end{subequations}
In the classical limit one achieves the equations of motion of OPA:
\begin{subequations}
\begin{gather}
\partial_z A_s\left(z\right)=-i\kappa e^{-i\Delta k z} A^*_i\left(z\right) A_p\left(z\right),\\
\partial_z A_i\left(z\right)=-i\kappa e^{-i\Delta k z} A^*_s\left(z\right) A_p\left(z\right),\\
\partial_z A_p\left(z\right)=-i\kappa e^{i\Delta k z} A_s\left(z\right) A_i\left(z\right),
\end{gather}
\label{eq: OPA_eq}
\end{subequations}

\subsection{SPDC in Different Directions}
\label{app: SPDC in Different Directions}
The process of SPDC can produce pairs of signal-idler photons in different directions. 
When the system is phase-matched in the same direction of the incident pump propagation (which is perpendicular to the crystal surface; call it the z direction) the output signal and idler waves of wavelengths $\lambda_s$ and $\lambda_i$ and propagation directions of $\theta_s$ and $\theta_i$, the best performance of the process will be when the process is phase-matched. The energy conservation and momentum conservation in the $\hat{X}-\hat{Y}$ plane are:
\begin{subequations}
    \begin{equation} \label{eq: energy conserv}
        \frac{1}{\lambda_s}+\frac{1}{\lambda_i}=\frac{1}{\lambda_p},
    \end{equation}
        \begin{equation} \label{eq: Y phase-matching}
        \frac{\sin{\theta_s}}{\lambda_s}=\frac{\sin{\theta_i}}{\lambda_i},
    \end{equation}
\end{subequations}
and the mismatch in such process is
\begin{equation}
  \Delta k= \Delta k\left(\lambda_s,\theta_s,T\right)- \Delta k\left(\lambda_s,0,T_{ppm}\right) ,
\end{equation}
where
\begin{equation}
\begin{split}
    \Delta k & \left(\lambda_s,\theta_s,T\right)=\frac{2\pi}{\lambda_p} n_T\left(\lambda_p,0\right)
    \\ & -\left(\frac{2\pi}{\lambda_s}  n_T\left(\lambda_s,\theta'_s\right) \cos{\theta'_s}+\frac{2\pi}{\lambda_i} n_T\left(\lambda_i,\theta'_i\right) \cos{\theta'_i}\right) ,
\end{split}
    \label{eq: QPM2}
\end{equation}
and the subscript ``$ppm$'' stands for perfect phase-matching which is under our assumptions, designed in the $\hat{Z}$ direction. $\theta'_j$ and $\theta_j$ satisfy Snell's law:
$n_T\left(\lambda_j,\theta'_j\right) \sin{\theta'_j}=\sin{\theta_j}$.

\subsection{Sensitivity of SPDC}
\label{app: Sensitivity of SPDC}

\begin{figure*}[htbp]
     \centering
    \begin{minipage}{1\linewidth}
    \raggedright (a) \hspace{240pt} (b)
    \end{minipage}
    
    \includegraphics[width=0.49\textwidth]{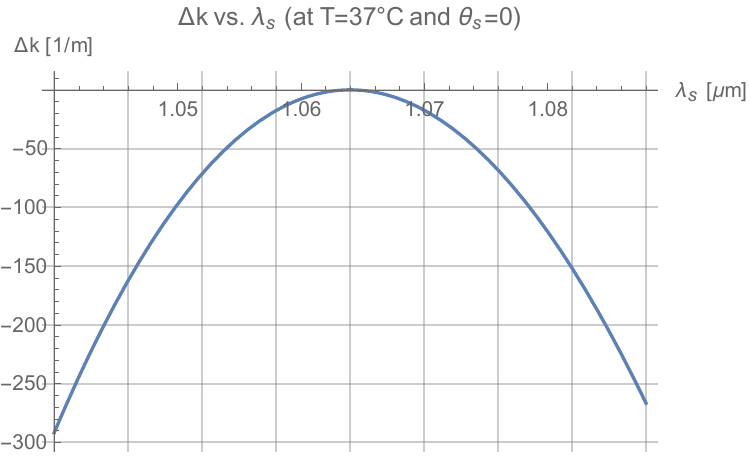}
    \includegraphics[width=0.49\textwidth]{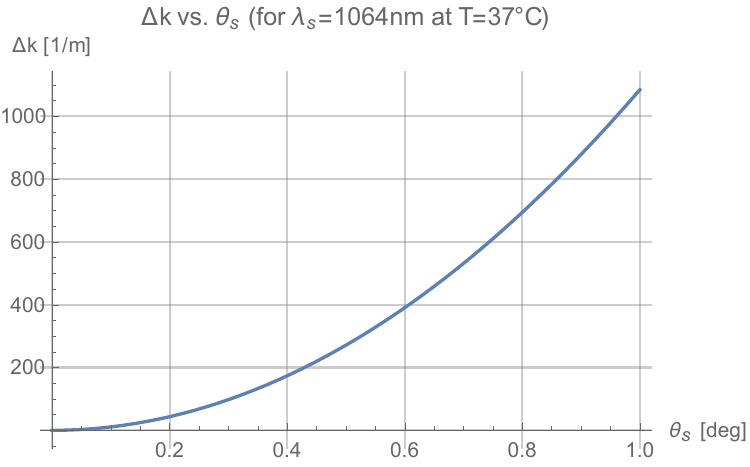}

    \begin{minipage}{1\linewidth}
    \raggedright (c) \hspace{240pt} (d)
    \end{minipage}
    
    \includegraphics[width=0.49\textwidth]{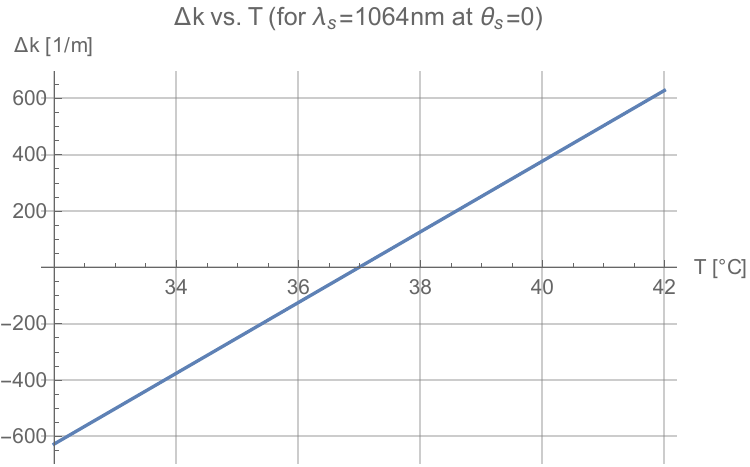}         \includegraphics[width=0.49\textwidth]{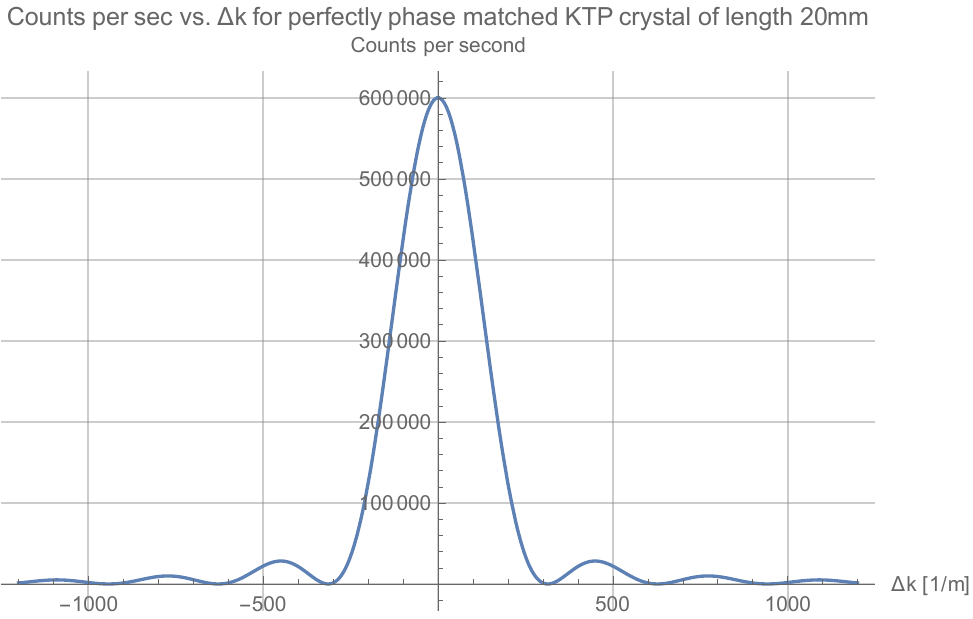}
    
    \begin{minipage}{1\linewidth}
    \raggedright (e)
    \end{minipage}
    
    \includegraphics[width=1\textwidth]{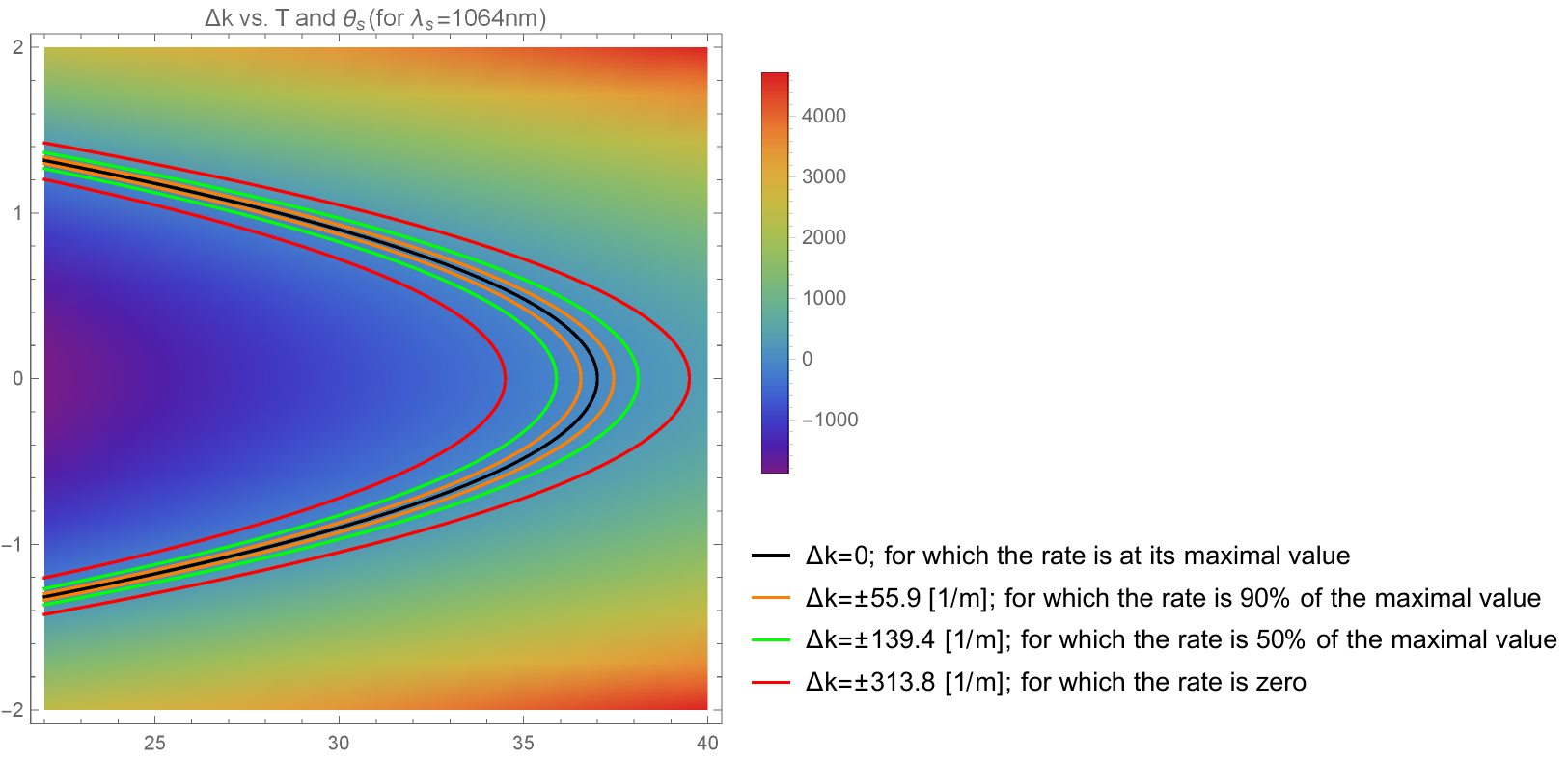}
    
     \caption{The sensitivity of $\Delta k$: we show in (a), (b), (c) how $\Delta k$ varies as each parameter changes while keeping the others at their optimal values, in (d) how the rate of generated entangles pairs change with $\Delta k$, and in (e) how $\Delta k$ varies as $T$ and $\theta_s$ change while keeping $\lambda_s=1064$nm.}
     \label{fig: S1 Sensitivity of DeltaK}
\end{figure*}

The SPDC, as a second order nonlinear process is very sensitive process, critically dependent on the parameters that influences the phase mismatch parameters.
The main obstacle to an efficient process in this case is the sensitivity of the phase mismatch to tiny variations of the signal/idler wavelength, the temperature, or even the angle of measurement. 
For example, in a perfectly quasi-phase-matched KTP crystal of length 20mm at $T=37^\circ C$, a deviation of only $0.445^\circ C$ in temperature (or 10.64nm in the signal's wavelength, or only $0.227^\circ$ in the measurement angle) is sufficient to lower the rate of generated photons to 90\% of the rate achieved by the error-free setup. 
This is due to the fact that in such processes $\Omega$ is $\sim 1 m^{-1}$, while $\Delta k$ changes with temperature by $\sim 125 {m^{-1\circ}} {C^{-1}}$. In this case, the system enters the harmonic regime and the rate of generated entangled photons starts to decrease significantly. 
For more details on the sensitivity of $\Delta k$ see Fig. \ref{fig: S1 Sensitivity of DeltaK} in which we show in (a), (b), (c) how $\Delta k$ varies as each parameter changes while keeping the others at their optimal values, in (d) how the rate of generated entangles pairs change with $\Delta k$, and in (e) how $\Delta k$ varies as $T$ and $\theta_s$ change while keeping $\lambda_s=1064$nm.

\subsection{Quantum $(u,v,w)$ Vector and Bloch Hyperboloid for SPDC}
\label{app: Quantum $(u,v,w)$ Vector and Bloch Hyperboloid for SPDC}

\begin{figure}[htbp]
         \centering
    \includegraphics[width=\linewidth]{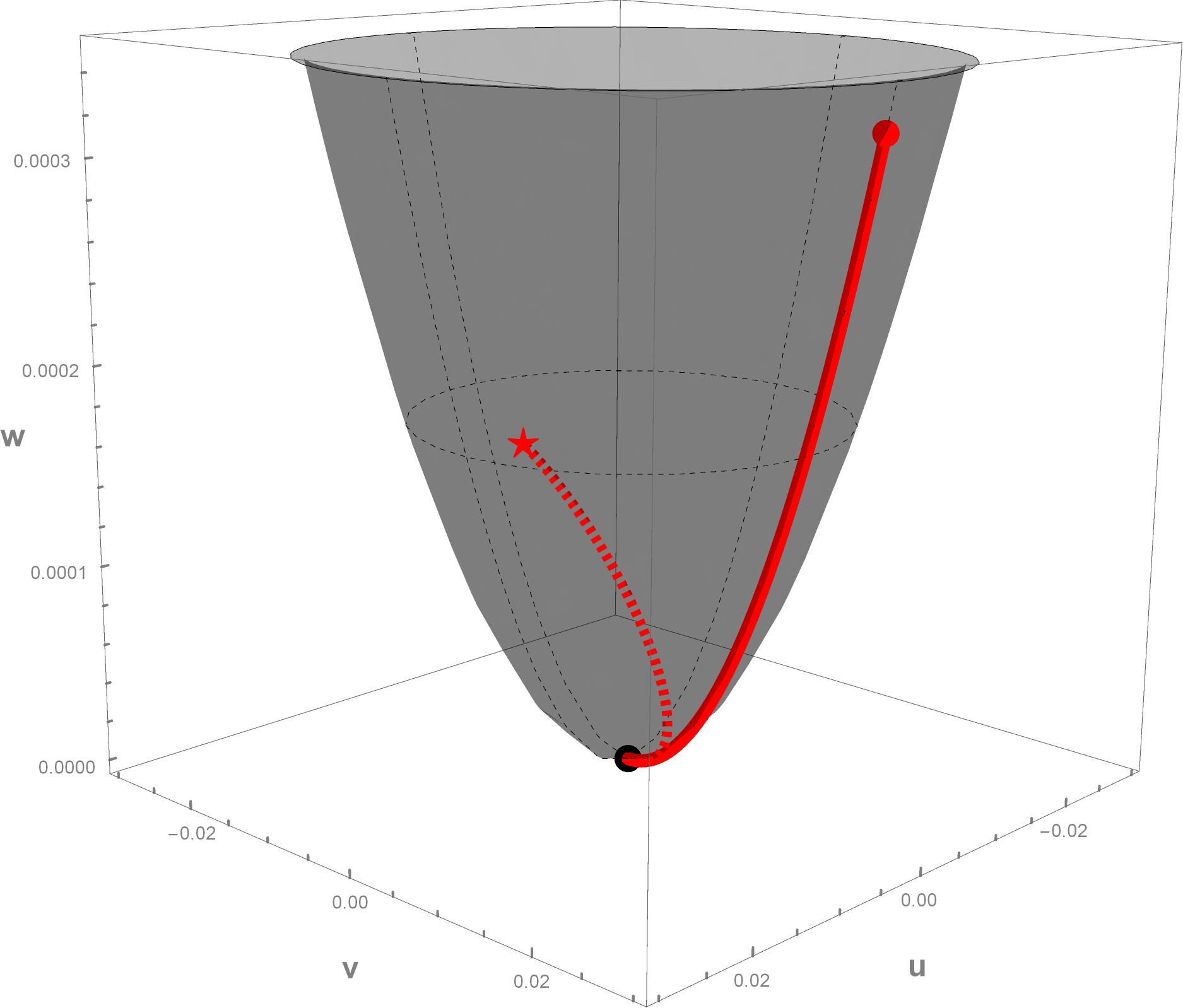}
     \caption{The $(u,v,w)$ trajectory in the 2 cm PP QPM crystal at the QPM temperature (the continuous line) and at 1 Celsius degree deviation from it (the dotted line). As one can see, the $w$ value at the end of the crystal is very sensitive to errors.}
     \label{fig: S2 PP trajectory}
\end{figure}

Like the Bloch sphere, on which lies the $(u, v, w)$ vector that can be defined from the fields that follow SU(2) dynamics, one can similarly define a $(u, v, w)$ vector for the case of SU(1,1), which govern SPDC and optical parametric amplification (OPA). For an SU(1,1) process, the operator vector $\left(\hat{u}, \hat{v}, \hat{w}\right)$ has an expectation value that is restricted to move on a hyperboloid. 
In the undepleted SPDC process, we define:
\begin{subequations}
\begin{gather}
    \hat{u}\left(z\right) = \hat{c}_s\left(z\right) \hat{c}_i\left(z\right) + \hat{c}^\dagger_s\left(z\right) \hat{c}^\dagger_i\left(z\right),\\
    \hat{v}\left(z\right) = i\left(\hat{c}_s\left(z\right) \hat{c}_i\left(z\right) - \hat{c}^\dagger_s\left(z\right) \hat{c}^\dagger_i\left(z\right)\right),\\
    \hat{w}\left(z\right) = \hat{c}^\dagger_s\left(z\right) \hat{c}_s\left(z\right) + \hat{c}_i\left(z\right) \hat{c}^\dagger_i\left(z\right),
\end{gather}
    \label{eq: uvw}
\end{subequations}
where $\hat{c}_j \equiv e^{i \left(\omega_j + \frac{\Delta k}{2}\right)z} \hat{a}_j$ (for $j=s,i$). Since $\hat{a}_s,\hat{a}^\dagger_i$, as operators, satisfy the same dynamics as $A_s,A^*_i$ in OPA (the undepleted-pump regime), as in the classical case, the vector $\left(\hat{u}, \hat{v}, \hat{w}\right)$ satisfies:
\begin{equation}
    \frac{d}{dz}
    \begin{pmatrix}
         \hat{u}\left(z\right)\\
         \hat{v}\left(z\right)\\
         \hat{w}\left(z\right)\\
    \end{pmatrix} =
    \begin{pmatrix}
         0 & \Delta k & 0\\
         -\Delta k & 0 & 2\Omega \\
         0 & 2\Omega & 0
    \end{pmatrix}
    \begin{pmatrix}
         \hat{u}\left(z\right)\\
         \hat{v}\left(z\right)\\
         \hat{w}\left(z\right)\\
    \end{pmatrix}.
    \label{eq: dynamics of uvw}
\end{equation}
Therefore, the expectation values $u\equiv\langle\hat{u}\rangle,v\equiv\langle\hat{v}\rangle,w\equiv\langle\hat{w}\rangle-1$ satisfy
\begin{equation}
    u^2+v^2-\left(w+1\right)^2=-1,
    \label{eq: hyperboloid}
\end{equation}
so that the vector $(u,v,w+1)$ moves on this hyperboloid during the process. As an example, we show in Fig. \ref{fig: S2 PP trajectory} the $(u,v,w)$ trajectory in a 2 cm PP QPM crystal at the QPM temperature and at 1 Celsius degree deviation from it. The sensitivity of the process to such error or deviation is seen via the $w$ value at the end of the crystal.

\section{Detuning Modulated Composite Segments for Robust SPDC and OPA Processes}
\label{app: Detuning Modulated Composite Segments for Robust SPDC and OPA Processes}

\subsection{Polings for The Designs and Manufacturing Limitations}
\label{app: Polings for The Designs and Manufacturing Limitations}

\begin{figure*}[htbp]
    \centering
    \includegraphics[width=1\textwidth,
    trim=0cm 2.5cm 0cm 2.5cm, clip]
    {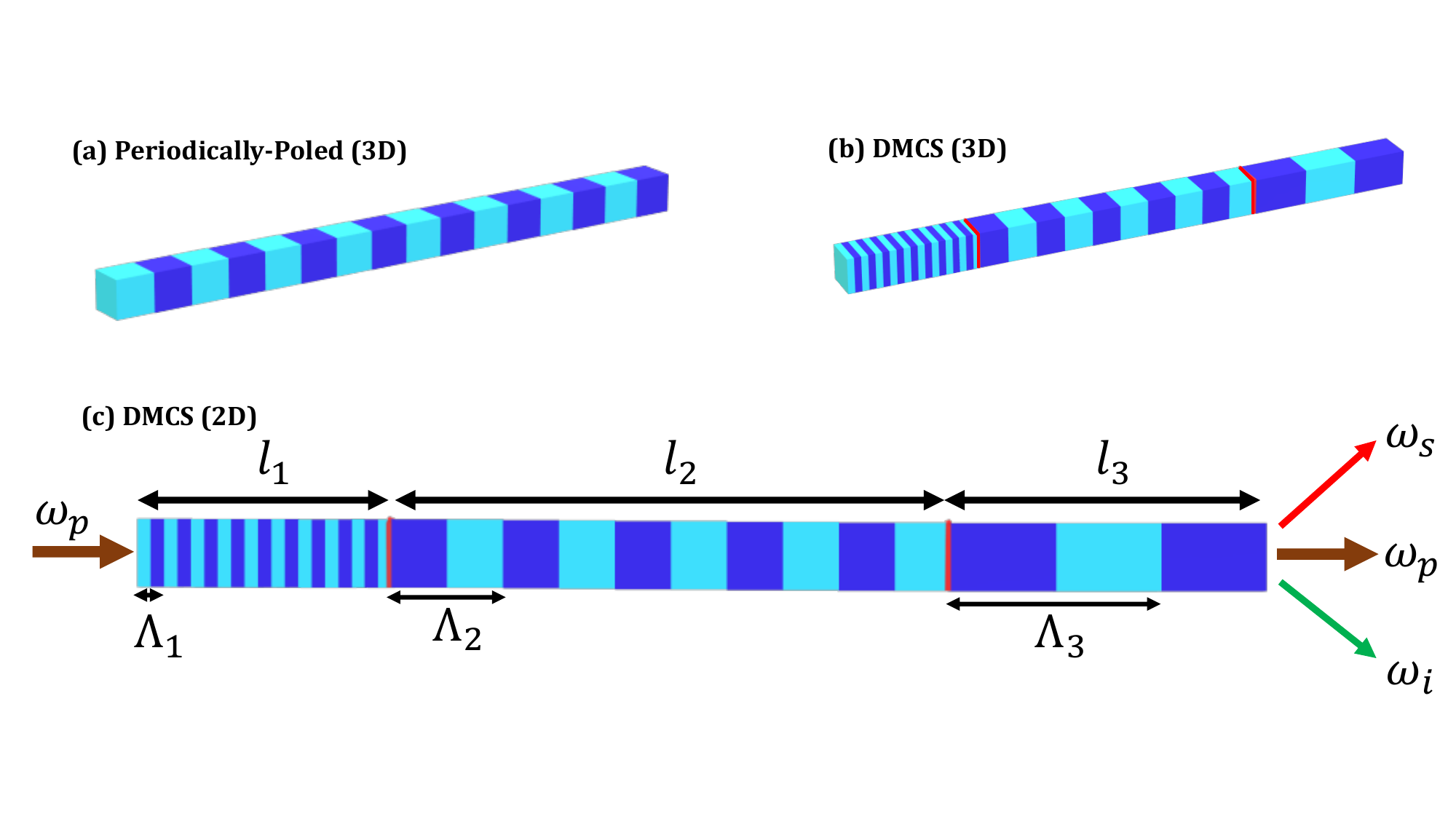}
    \caption{A schematic illustration of the poling design of a composite crystal compared to the periodically poled and perfectly quasi-phase-matched one.}
    \label{fig: S3 DMCS crystal}
\end{figure*}

The periodically poled scheme for a crystal with a length of $L$ and a phase-mismatch of $\Delta k$ that matches a process of $\lambda_p \rightarrow \lambda_s + \lambda_i$ at temperature $T_p$ is designed with the following poling (provided $L\gg \frac{2\pi}{\Delta k}$):
\begin{equation}
\begin{split}
    \chi^{\left(2\right)} \left(z\right) & = \lvert \chi^{\left(2\right)} \rvert \mathrm{sign}\left( \cos{\left( \Delta k_{pol} \cdot z \right)} \right) 
    \\ & = \lvert \chi^{\left(2\right)} \rvert \mathrm{sign}\left( \cos{\left( \frac{2\pi}{\Lambda} \cdot z \right)} \right) ,
\end{split}
\end{equation}
where $\Lambda$ here is the the poling length or period. 
Illustrations of the periodic poling method are given in Fig. \ref{fig: S3 DMCS crystal}.

However, the manufacturing usually restricts the lengths of domains with the same sign of the nonlinear susceptibility to be a multiple of some $\delta l$, and to be larger than some $\frac{\Lambda_{\mathrm{min}}}{2}$. 
So, in order to achieve some $\Delta k$ of the crystal, we use the following method for the sign of the nonlinear susceptibility:
\begin{equation}
    \chi^{\left(2\right)} \left(z \in \left( (j-1)\delta l , j \delta l \right) \right)= \lvert \chi^{\left(2\right)} \rvert \mathrm{sign}\left( \cos {\left( \Delta k_{pol} \cdot j \delta l \right)} \right) ,
    \label{eq: Poling}
\end{equation}
where $j$ runs from $1$ to $\frac{L}{\delta l}$. If we get continuous lengths less than $\frac{\Lambda_{\mathrm{min}}}{2}$ this means that we cannot design such $\Delta k$ for the crystal. Typical values as we know up-to-day are $\delta l=25$nm and $\Lambda_{\mathrm{min}}=3000$nm. For a segmented design, where $\Delta k_{pol}$ is a function of $z$, Eq. (\ref{eq: Poling}) becomes:
\begin{equation}
\begin{split}
    \chi^{\left(2\right)} & \left(z \in \left( (j-1)\delta l , j \delta l \right) \right)
    \\ & = \lvert \chi^{\left(2\right)} \rvert \mathrm{sign}\left[ \cos {\left( \int_{0}^{j \delta l} \Delta k_{pol} \left( z' \right) \,dz' \right)} \right] .
\end{split}
    \label{eq: Poling for segmented}
\end{equation}

\subsection{Selected Results}
\label{app: Selected Results}

Here we show eight selected designs of segmented KTP crystals designed for robust SPDC of $60$mW incident $532$nm pump with beam radius of $150\mathrm{\mu m}$ at $T_p=37^\circ C$. The signal and idler are of 1064nm wavelengths and at $\theta=0$ of each. The first six designs are all designs of anti-symmetric six segments:
\begin{equation}
    \Delta k_6=-\Delta k_1 \: , \: \Delta k_5=-\Delta k_2 \: , \: \Delta k_4=-\Delta k_3 \: ,
    \label{eq: antisymmetry}
\end{equation}
and the length of every segment with a half phase mismatch of $\delta$ is:
\begin{equation}
    l_j=\frac{\pi}{2\sqrt{\left(\frac{\Delta k_j}{2}\right)^2-\Omega^2}} .
    \label{eq: lengths}
\end{equation}
We chose such assumptions due to the analytical simplicity of finding robust designs. We present also another two selected designs of ten segments with the following $\Delta k$'s vector:
\begin{equation}
\begin{split}
    \left( \right. & \: \Delta k_1 \: , \: -\Delta k_1 \: , \: \Delta k_2 \: , \: -\Delta k_2 \: , \: \Delta k_3 \: 
    \\ & \left. , \: -\Delta k_3 \: , \: \Delta k_2 \: , \: -\Delta k_2 \: , \: \Delta k_1 \: , \: -\Delta k_1 \: \right) \: ,
\end{split}
    \label{eq: antisymmetry10}
\end{equation}
and with lengths as in Eq. \ref{eq: lengths}.

The eight designs are given in Table \ref{table: DMCS designs}, in which we specify for each crystal the design of the 3 representative segments, the total length of the crystal, the efficiency of generating entangles pairs compared to the one generated by a perfectly phase-matched crystal with the same total length, and the robustness width, which is the width (in the temperature axis) in which the rate of generated pairs of entangled photons decreases to 90\% of its maximum.

\begin{table}[htbp]
\centering
\resizebox{\linewidth}{!}{
  \begin{tabular}{|c|c||c|c|c|}
 \hline
 \multicolumn{5}{|c|}{DMCS designs of robust SPDC process at $T=37^\circ C$} \\
 \hline
 design & $\Delta k_1,\Delta k_2,\Delta k_3$ [1/m] & $L_{\mathrm{total}}$ [mm] & $\frac{\mu}{\mu_{\mathrm{ppm}}}$ & robustness width [$^\circ$C] \\
 \hline
 \hline
 I & {3935.29,-3935.29,-449.769} & 17.163 & 15.978\% & 11.5155 \\
 II & {8432.76,-8432.76,-887.712} & 8.568 & 17.237\% & 22.1545 \\
 III & {1840,445,-9700} & 9.091 & 15.800\% & 21.092 \\
 IV & {4000,54800,-886} & 8.777 & 16.708\% & 16.057 \\
 V & {3985,1000,-16250} & 8.247 & 15.459\% & 23.482 \\
 VI & {3563.45,35634.5,-779.505} & 10 & 16.984\% & 14.838 \\
 \hline
 VII & {34482.8,12069,-862.069} & 8.694 & 18.556\% & 20.740 \\
 VIII & {10285.7,67142.9,-857.143} & 8.739 & 18.607\% & 20.671 \\
 \hline
  \end{tabular}}
      \caption{The designs of the DMCS crystals: We show for each crystal the design of the 3 representative segments of phase mismatches of $\Delta k_{1,2,3}$ (see Eqs. (\ref{eq: antisymmetry},\ref{eq: lengths}) for (I-VI) and Eqs. (\ref{eq: antisymmetry10},\ref{eq: lengths}) for (VII-VIII)), the total length of the crystal, the efficiency of generating entangles pairs compared to the one generated by a perfectly phase-matched crystal with the same total length, and the robustness width, which is the width (in the temperature axis) in which the rate of generated pairs of entangled photons decreases to 90\% of its maximum.}
  \label{table: DMCS designs}
\end{table}

\begin{figure*}[htbp]
     \centering

    \begin{minipage}{1\linewidth}
    \raggedright (a) \hspace{240pt} (b)
    \end{minipage}

    \includegraphics[width=0.49\textwidth]{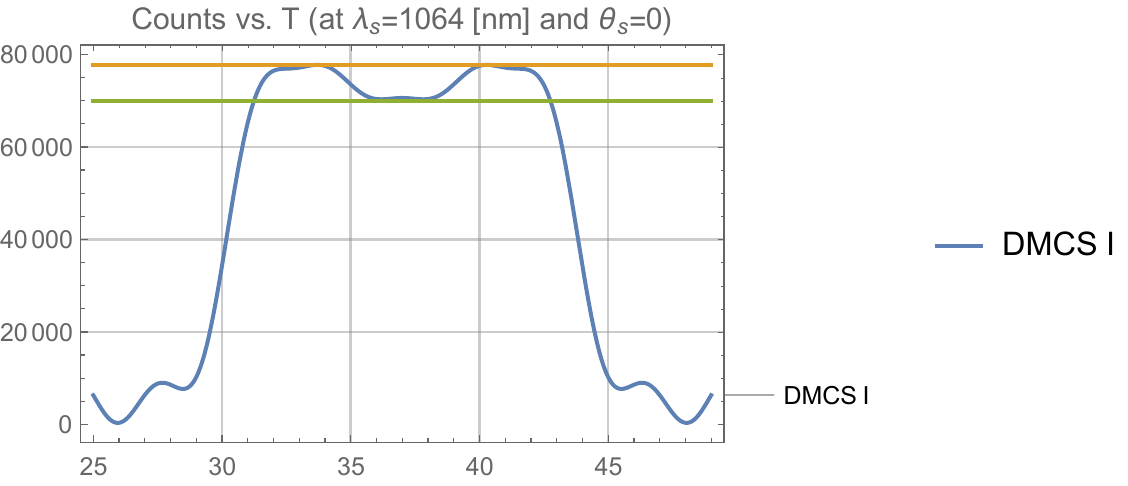}
    \includegraphics[width=0.49\textwidth]{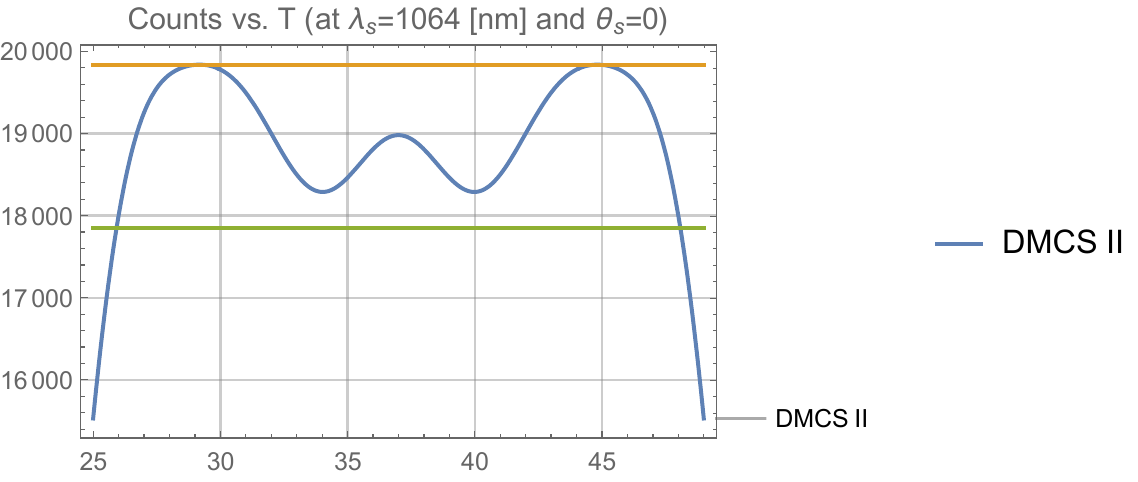}

    \begin{minipage}{1\linewidth}
    \raggedright (c) \hspace{240pt} (d)
    \end{minipage}

    \includegraphics[width=0.49\textwidth]{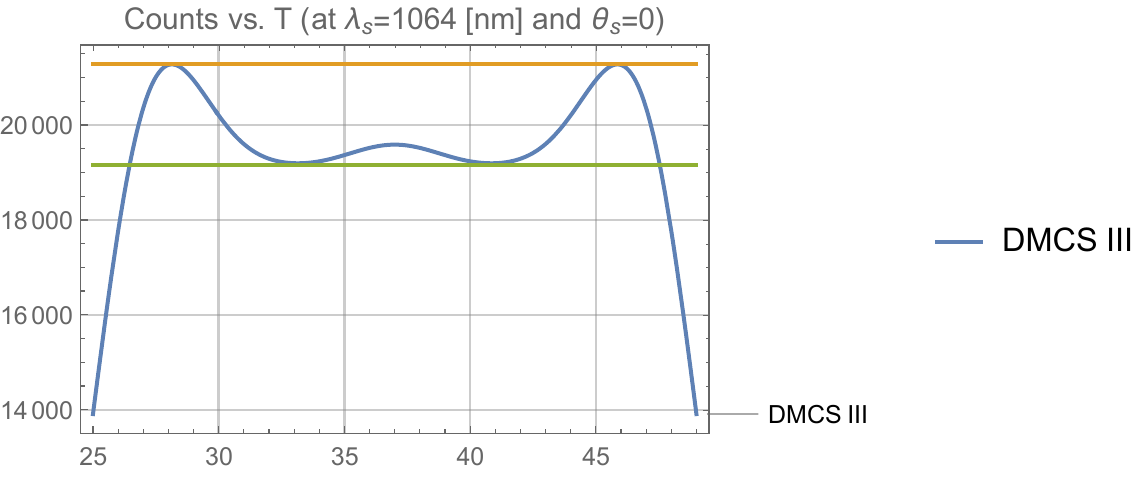}
    \includegraphics[width=0.49\textwidth]{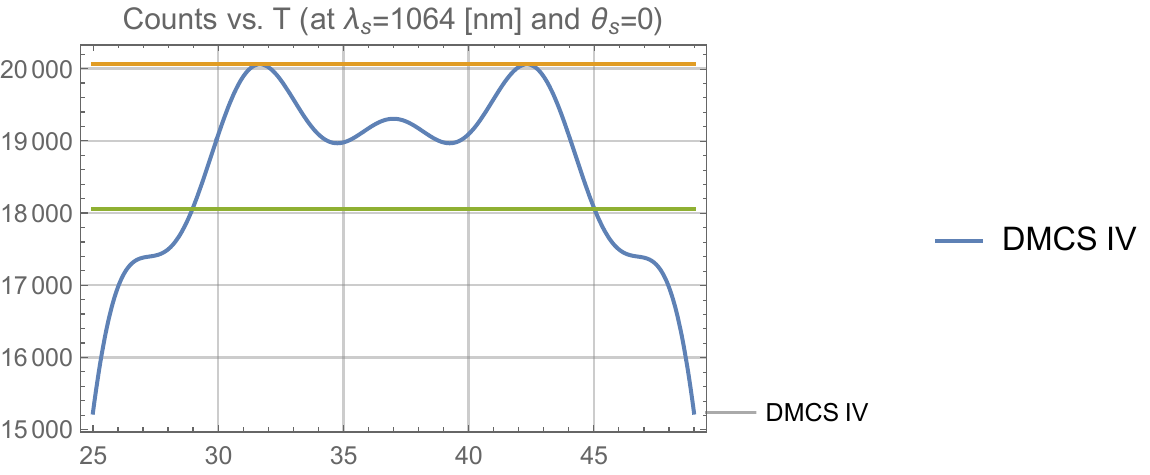}

    \begin{minipage}{1\linewidth}
    \raggedright (e) \hspace{240pt} (f)
    \end{minipage}

    \includegraphics[width=0.49\textwidth]{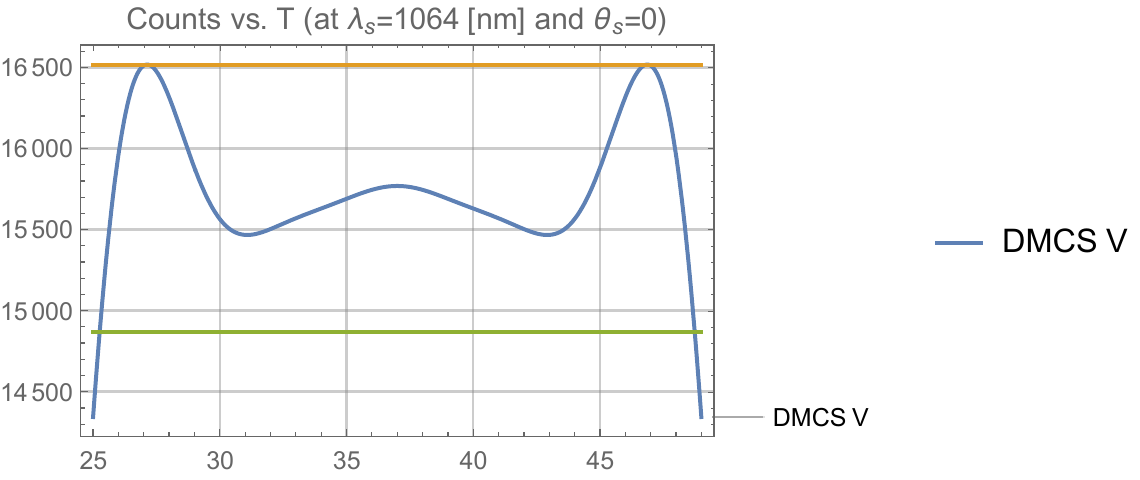}
    \includegraphics[width=0.49\textwidth]{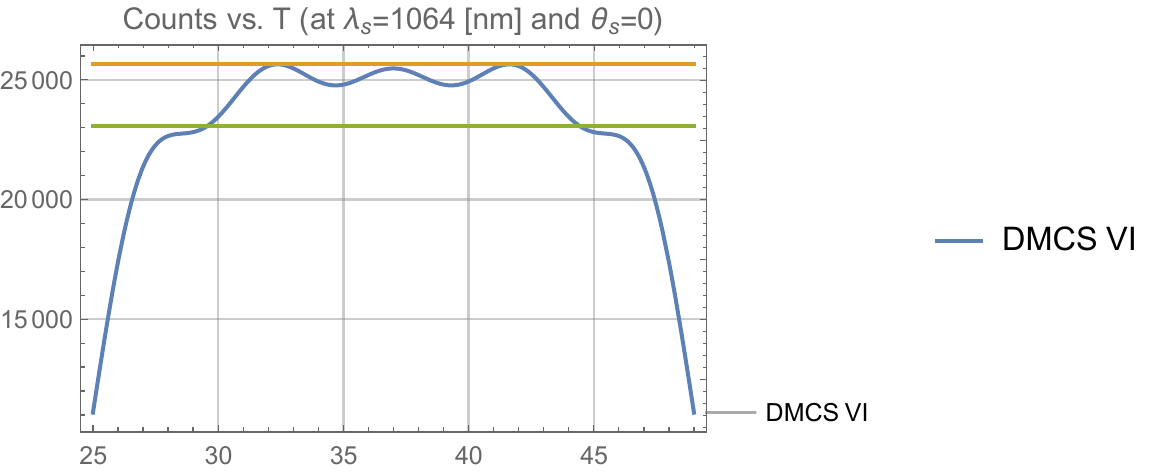}

    \begin{minipage}{1\linewidth}
    \raggedright (g) \hspace{240pt} (h)
    \end{minipage}

    \includegraphics[width=0.49\textwidth]{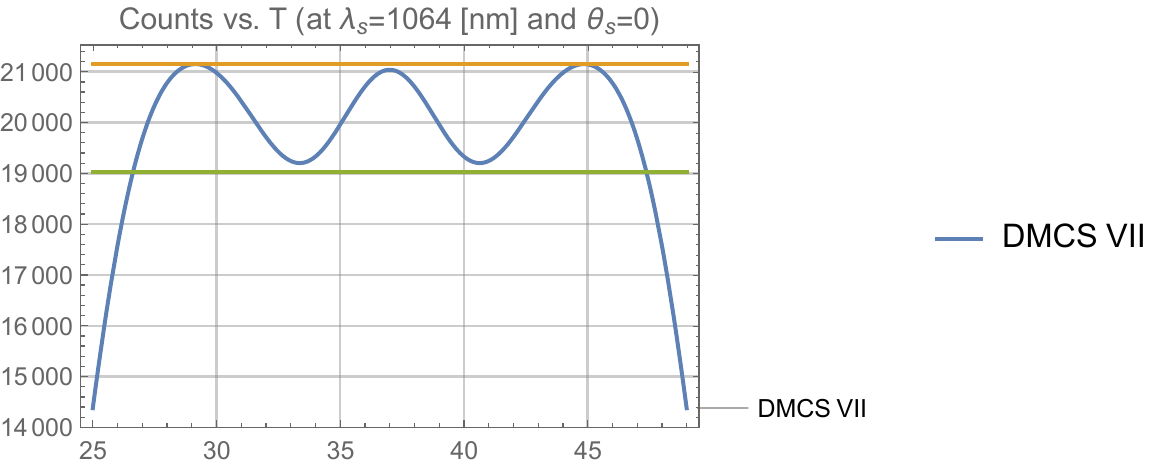}
    \includegraphics[width=0.49\textwidth]{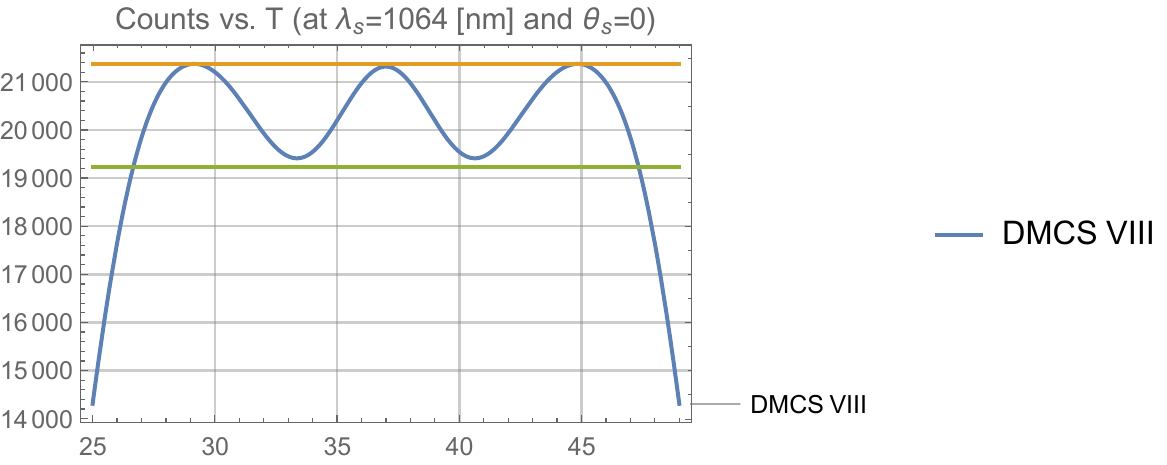}

     \caption{The number of generated entangled pairs per second as a function of the temperature for each DMCS design. In order to observe the robustness of the process, we added the two horizontal lines which are the lines of the maximum values and 90\% of them.}
     \label{fig: S4 counts vs T}
\end{figure*}

\begin{figure*}[htbp]
     \centering

    \begin{minipage}{1\linewidth}
    \raggedright (a) \hspace{240pt} (b)
    \end{minipage}

    \includegraphics[width=0.49\textwidth]{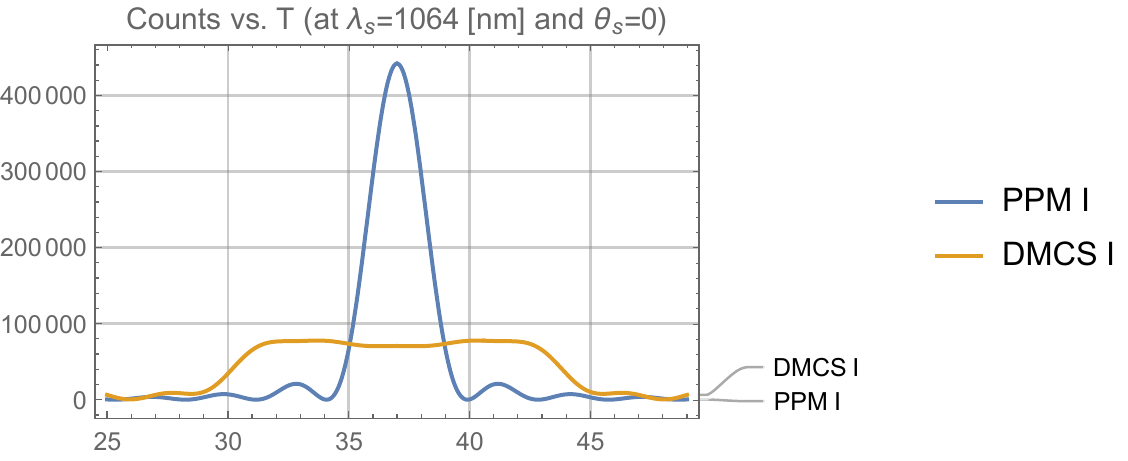}
    \includegraphics[width=0.49\textwidth]{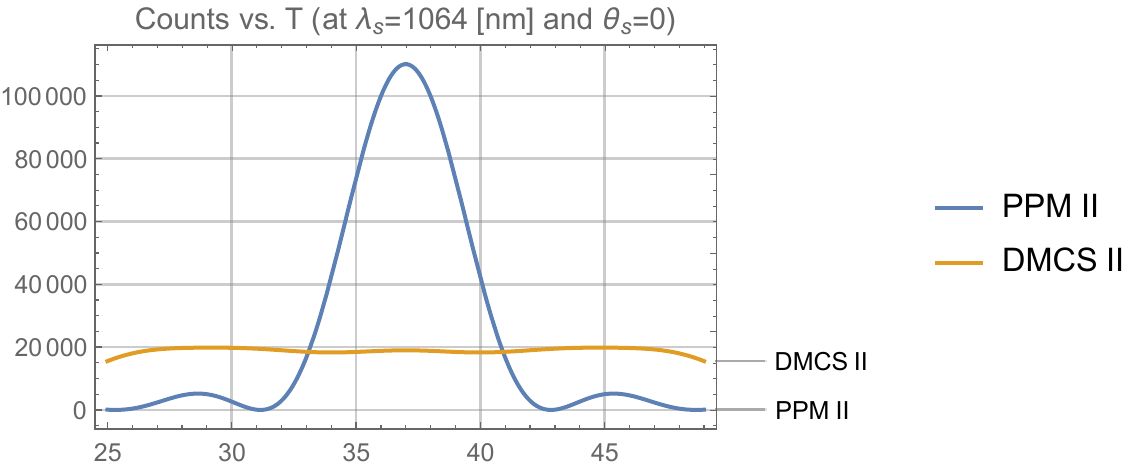}

    \begin{minipage}{1\linewidth}
    \raggedright (c) \hspace{240pt} (d)
    \end{minipage}

    \includegraphics[width=0.49\textwidth]{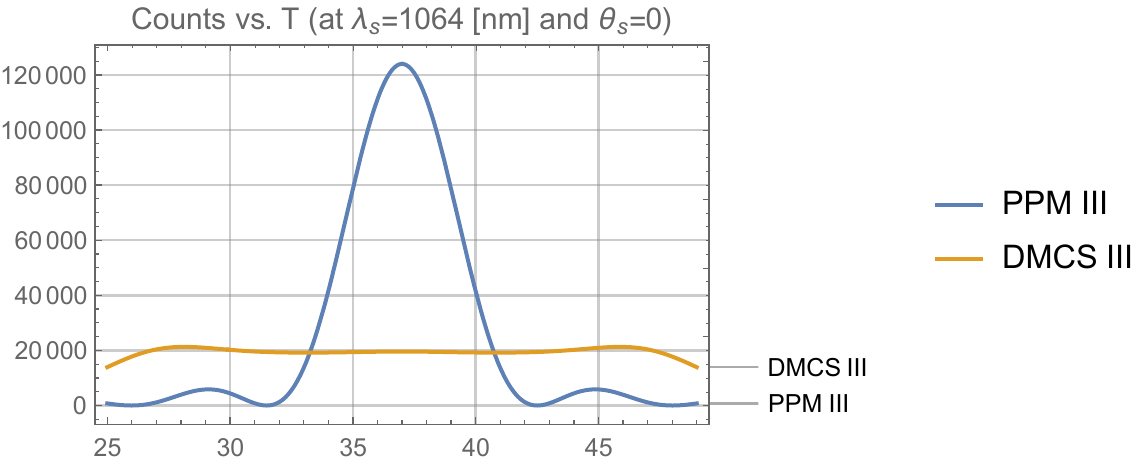}
    \includegraphics[width=0.49\textwidth]{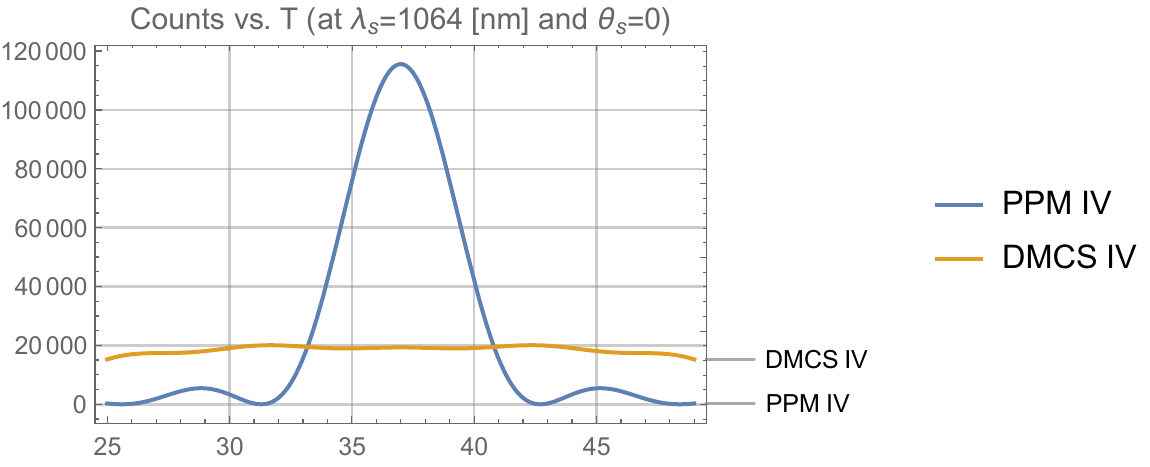}

    \begin{minipage}{1\linewidth}
    \raggedright (e) \hspace{240pt} (f)
    \end{minipage}

    \includegraphics[width=0.49\textwidth]{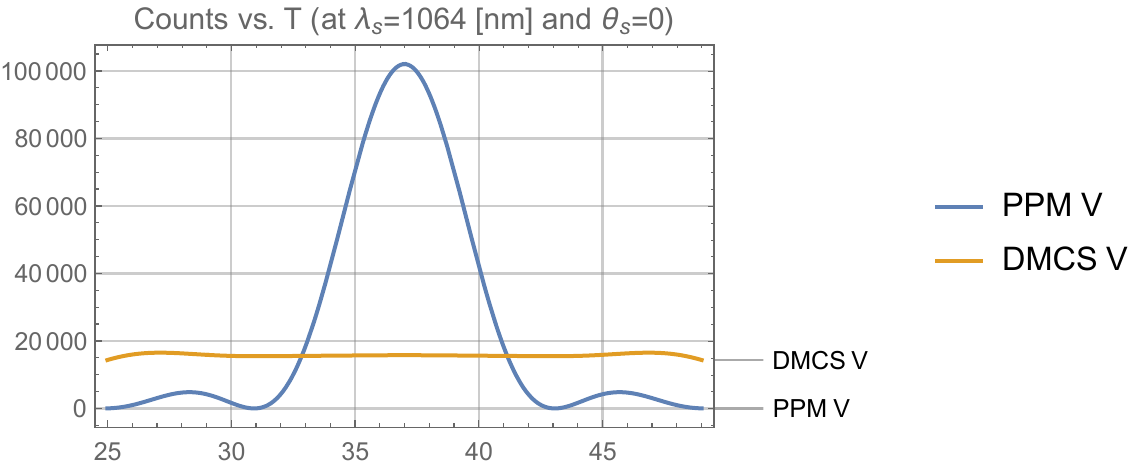}
    \includegraphics[width=0.49\textwidth]{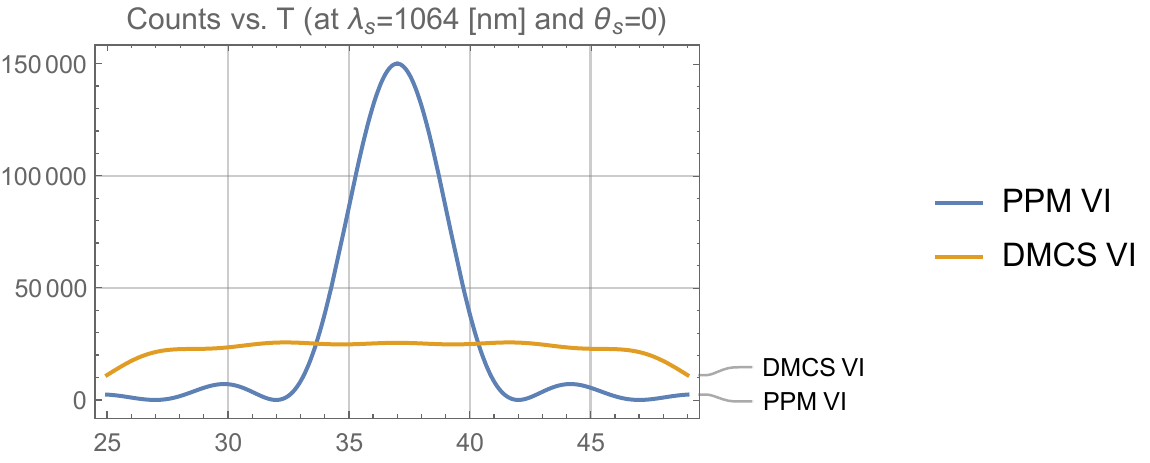}

    \begin{minipage}{1\linewidth}
    \raggedright (g) \hspace{240pt} (h)
    \end{minipage}

    \includegraphics[width=0.49\textwidth]{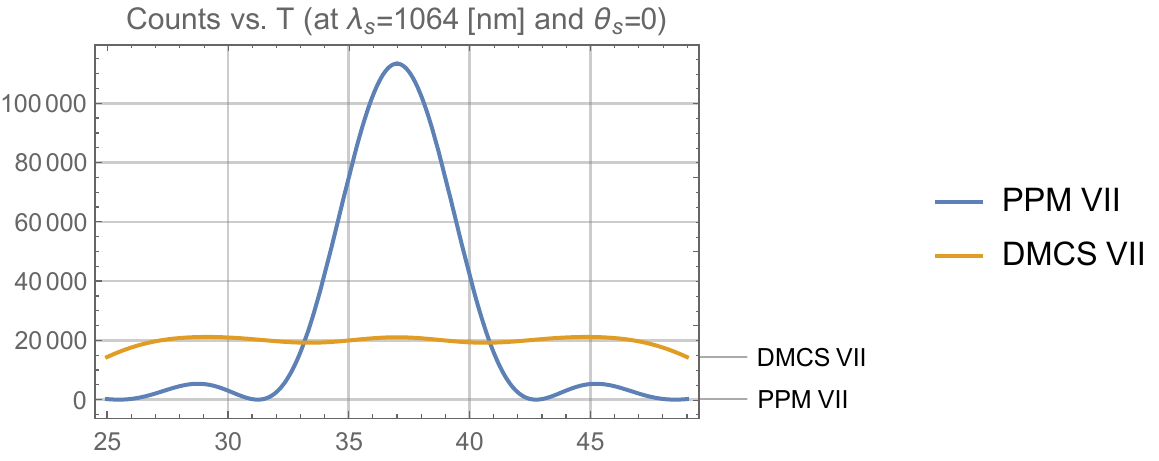}
    \includegraphics[width=0.49\textwidth]{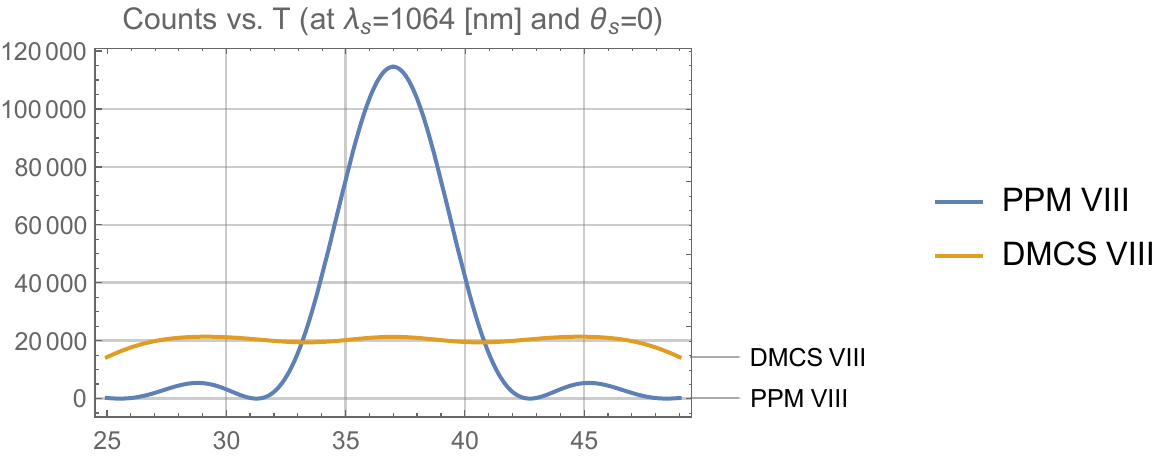}
    
     \caption{The number of generated entangled pairs per second as a function of the temperature for each DMCS design compared to the perfectly phase matched crystal with the same total length.}
     \label{fig: S5 counts vs T + PPM}
\end{figure*}

\begin{figure*}[htbp]
     \centering

    \begin{minipage}{1\linewidth}
    \raggedright (a) \hspace{156pt} (b) \hspace{156pt} (c)
    \end{minipage}

    \includegraphics[width=0.33\textwidth]{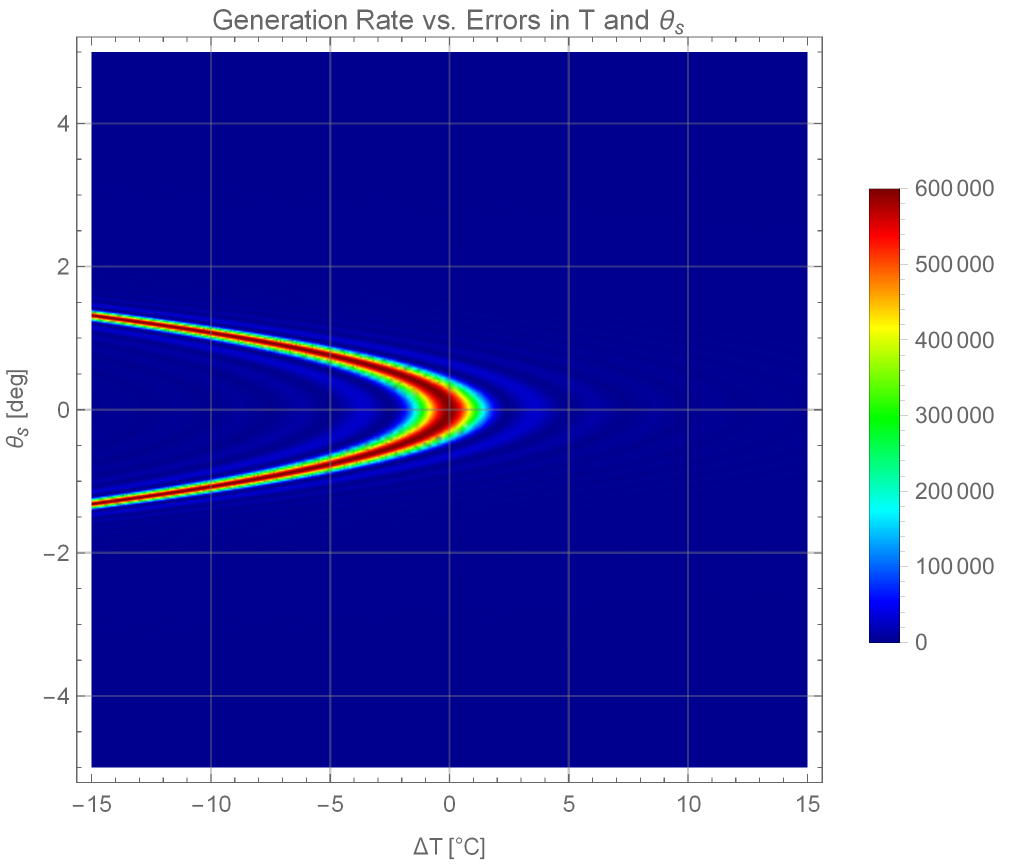}
    \includegraphics[width=0.33\textwidth]{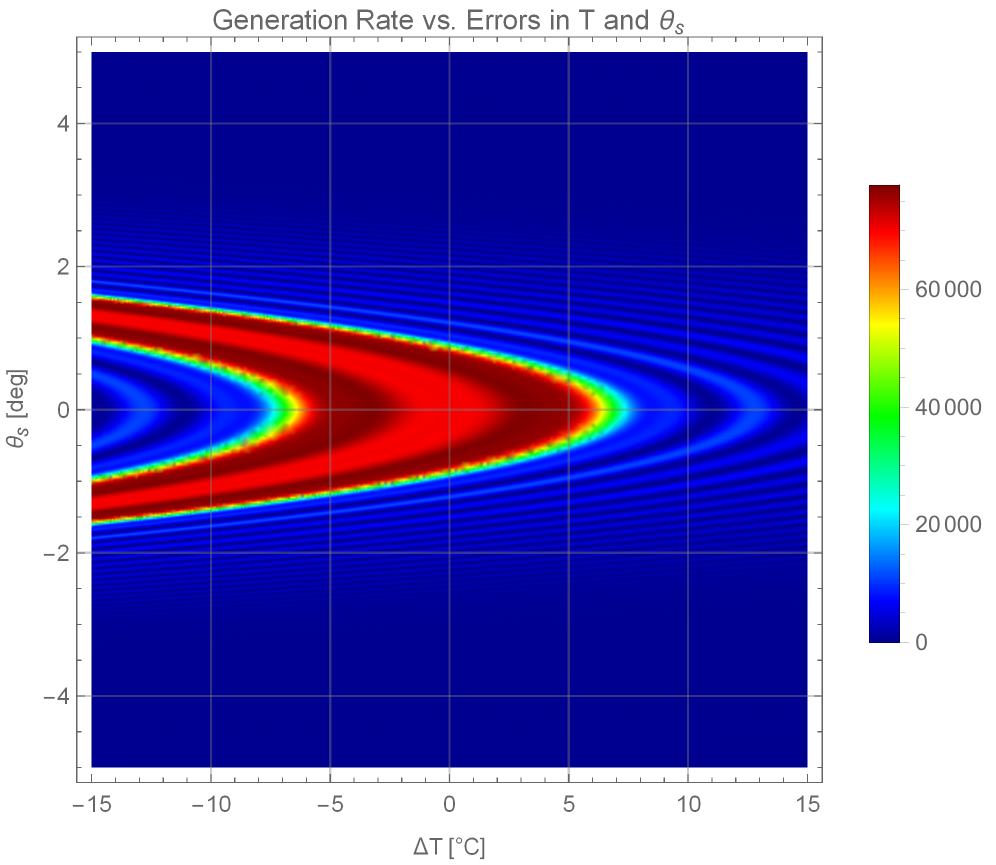}
    \includegraphics[width=0.33\textwidth]{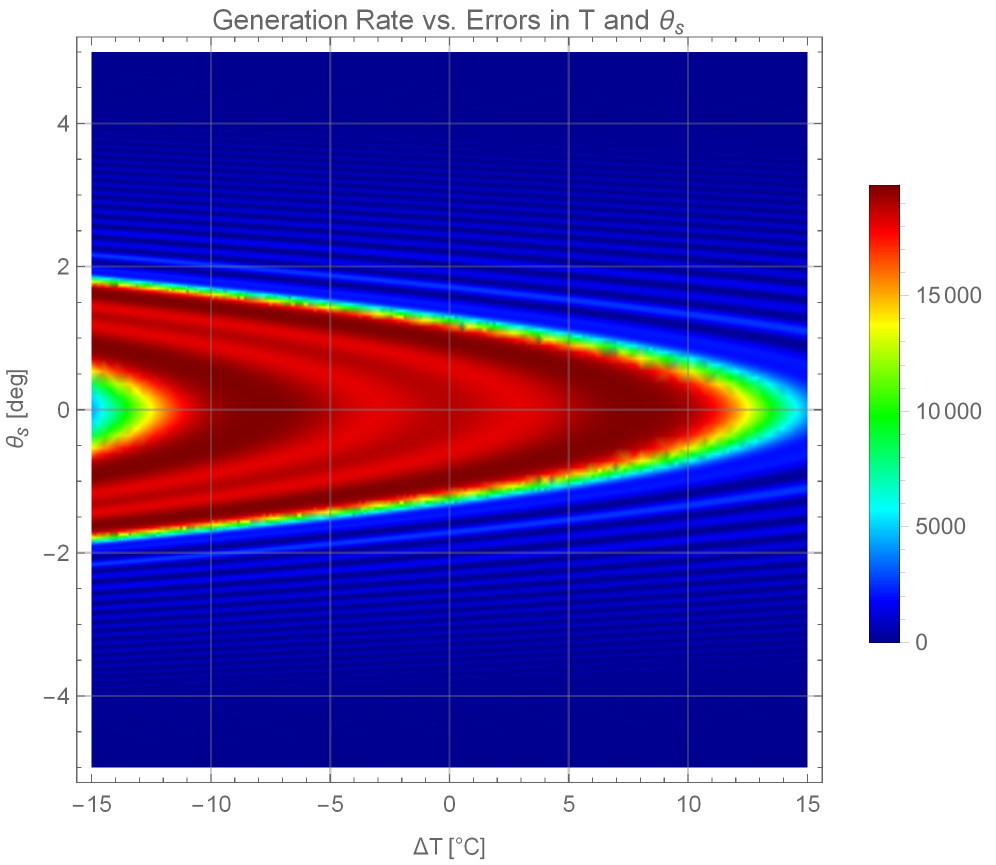}

    \begin{minipage}{1\linewidth}
    \raggedright (d) \hspace{156pt} (e) \hspace{156pt} (f)
    \end{minipage}

    \includegraphics[width=0.33\textwidth]{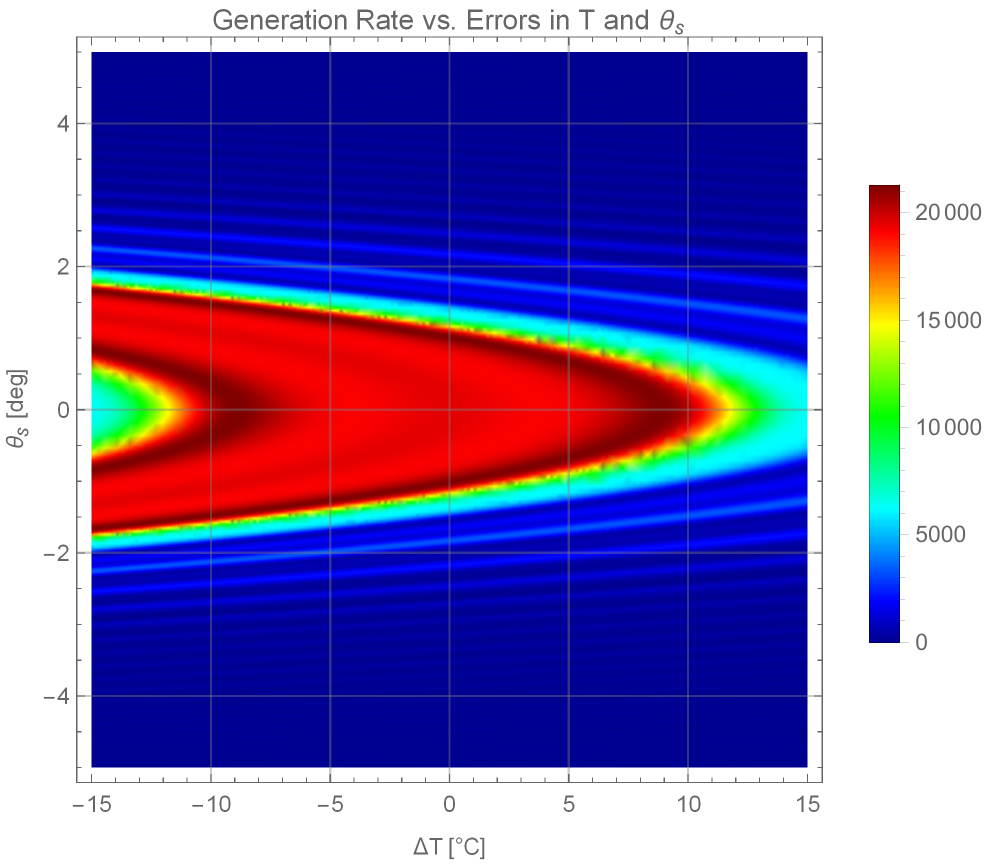}
    \includegraphics[width=0.33\textwidth]{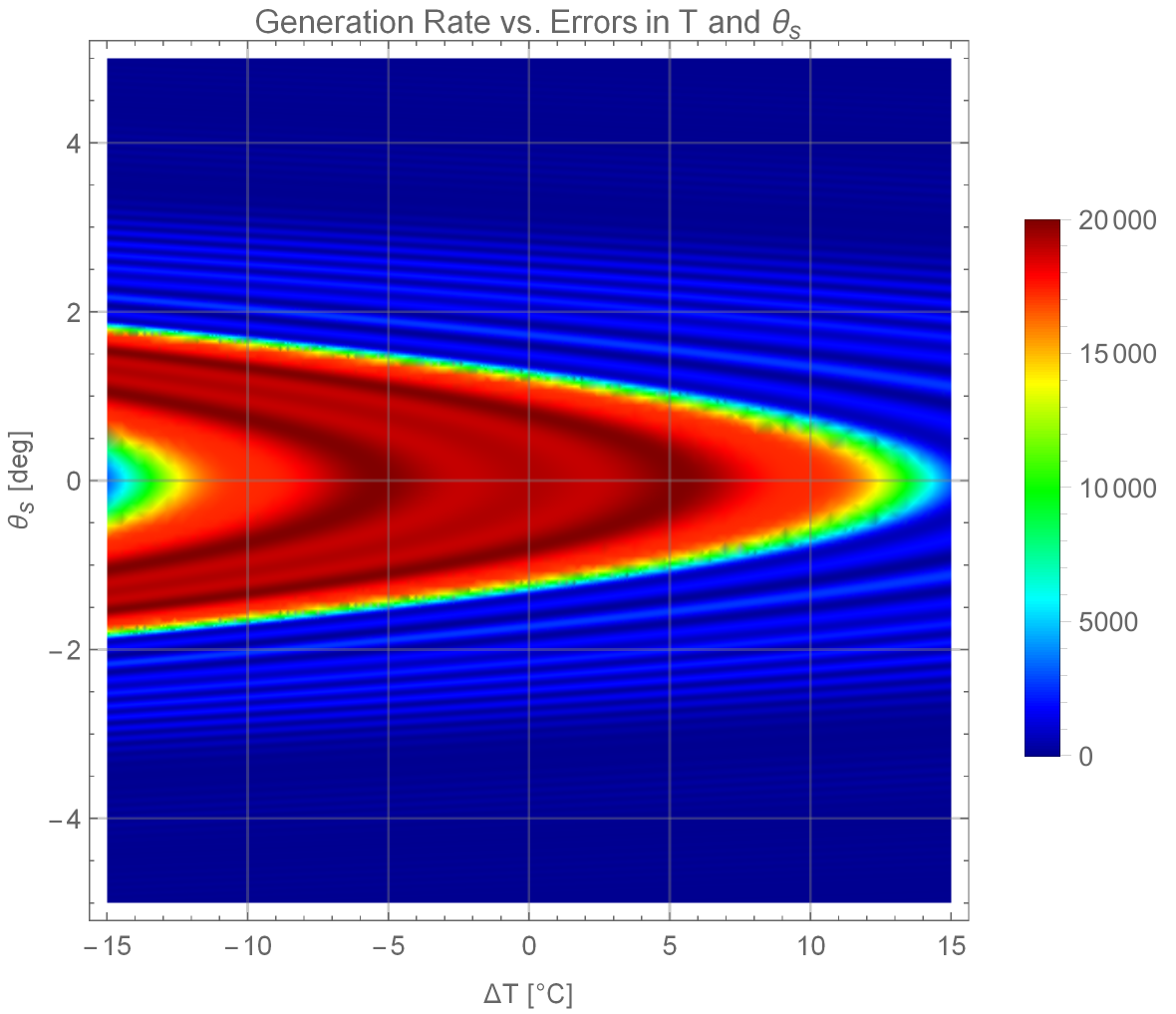}
    \includegraphics[width=0.33\textwidth]{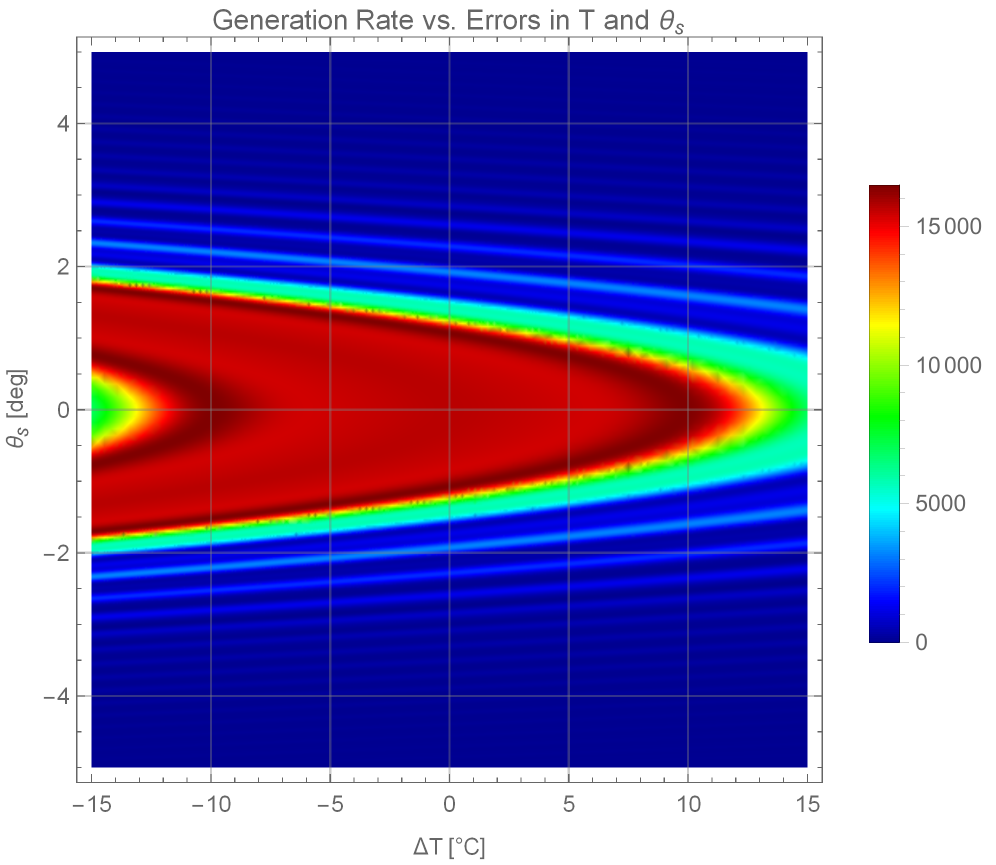}

    \begin{minipage}{1\linewidth}
    \raggedright (g) \hspace{156pt} (h) \hspace{156pt} (i)
    \end{minipage}

    \includegraphics[width=0.33\textwidth]{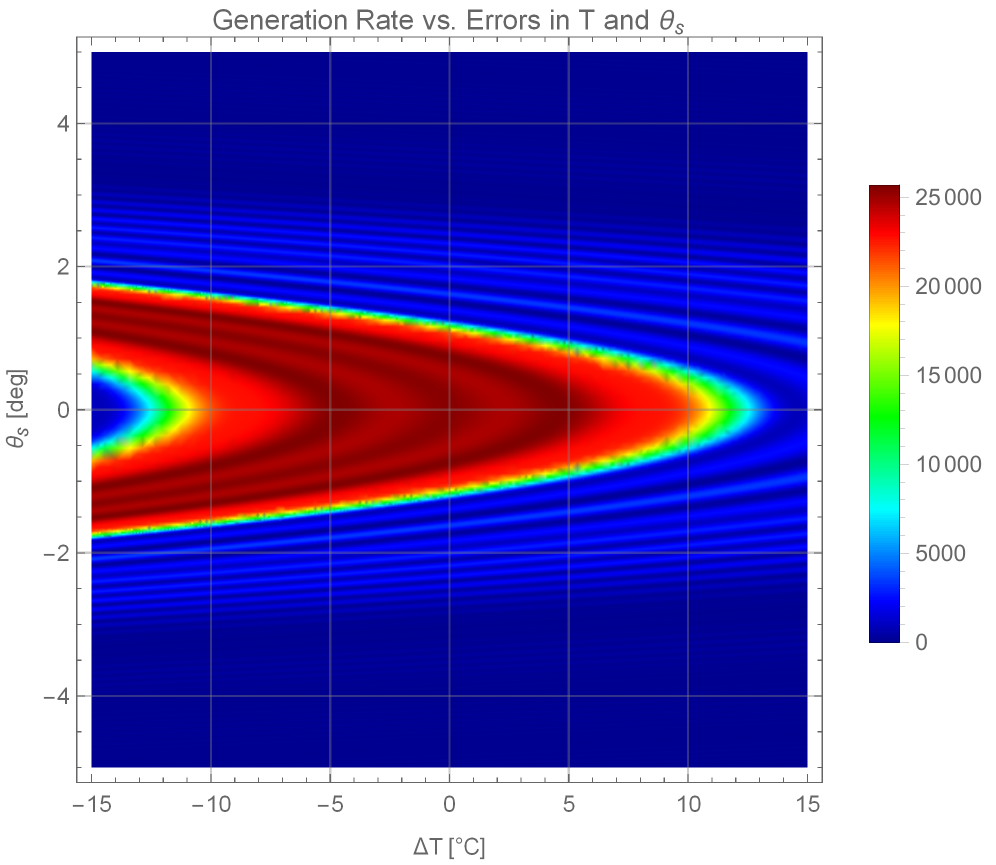}
    \includegraphics[width=0.33\textwidth]{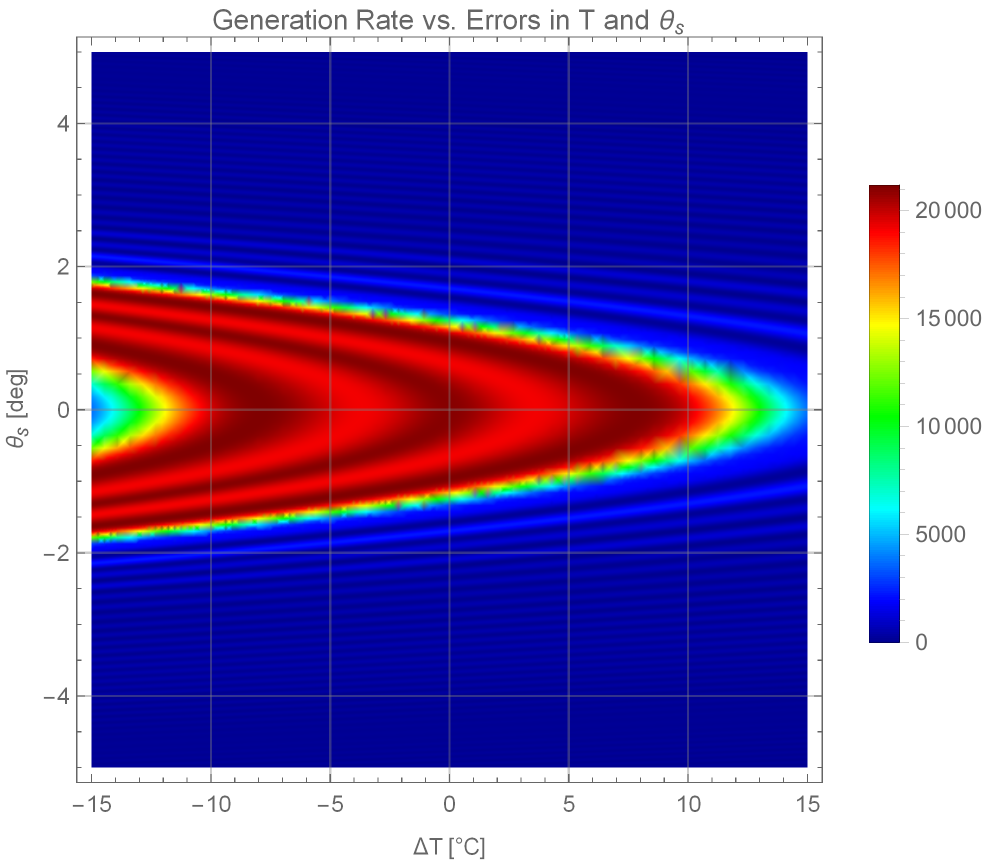}
    \includegraphics[width=0.33\textwidth]{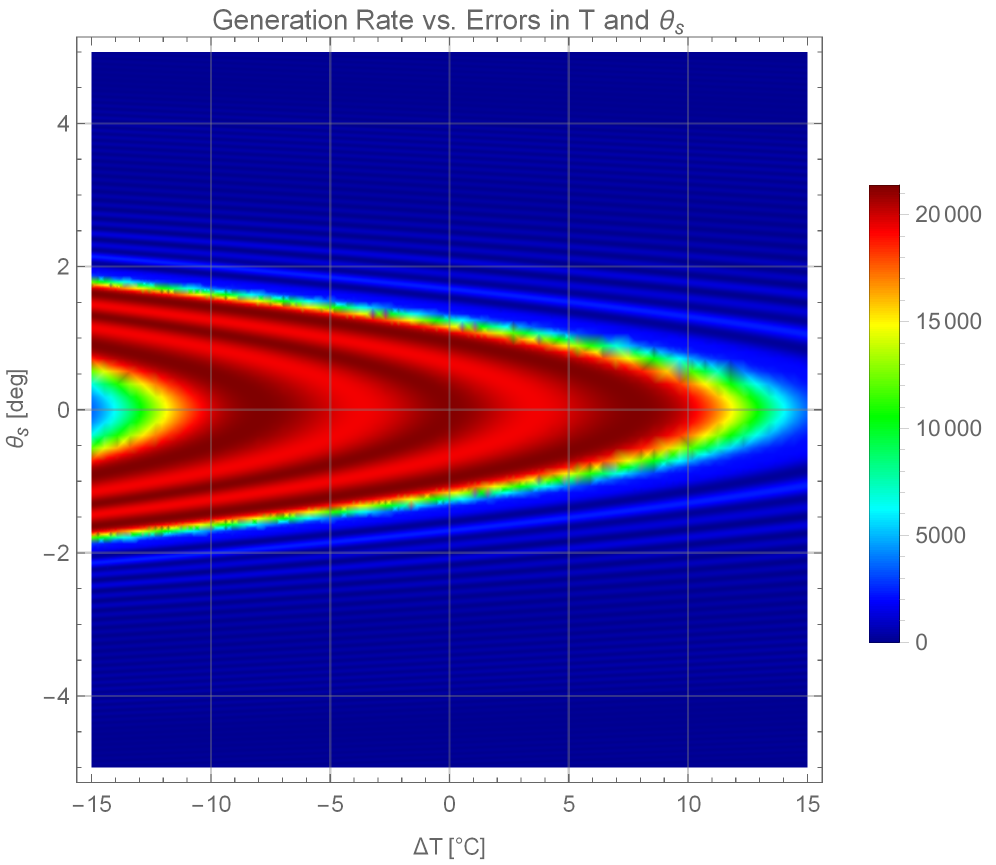}

     \caption{The number of generated entangled pairs per second as a function of the temperature deviation from $37^\circ C$ and $\theta_s$ for: (a) a perfectly phase matched crystal of length 20mm, (b,c,d,e,f,g,h,i) the DMCS crystals I,II,III,IV,V,VI,VII,VIII respectively.}
     \label{fig: S6 counts vs T and theta}
\end{figure*}

We show in Fig. \ref{fig: S4 counts vs T} the number of generated entangled pairs per second as a function of the temperature for each DMCS design, and in Fig. \ref{fig: S5 counts vs T + PPM} we show this compared to the perfectly phase matched crystal with the same total length. We also show in Fig. \ref{fig: S6 counts vs T and theta} the counts as a function of the temperature deviation from $37^\circ C$ and $\theta_s$ for a perfectly phase matched crystal of length 20mm and for our DMCS designs.

\subsection{Errors in The Pump's Intensity}
\label{app: Errors in The Pump's Intensity}

\begin{figure*}[htbp]
     \centering

    \begin{minipage}{1\linewidth}
    \raggedright (a) \hspace{156pt} (b) \hspace{156pt} (c)
    \end{minipage}

    \includegraphics[width=0.33\textwidth]{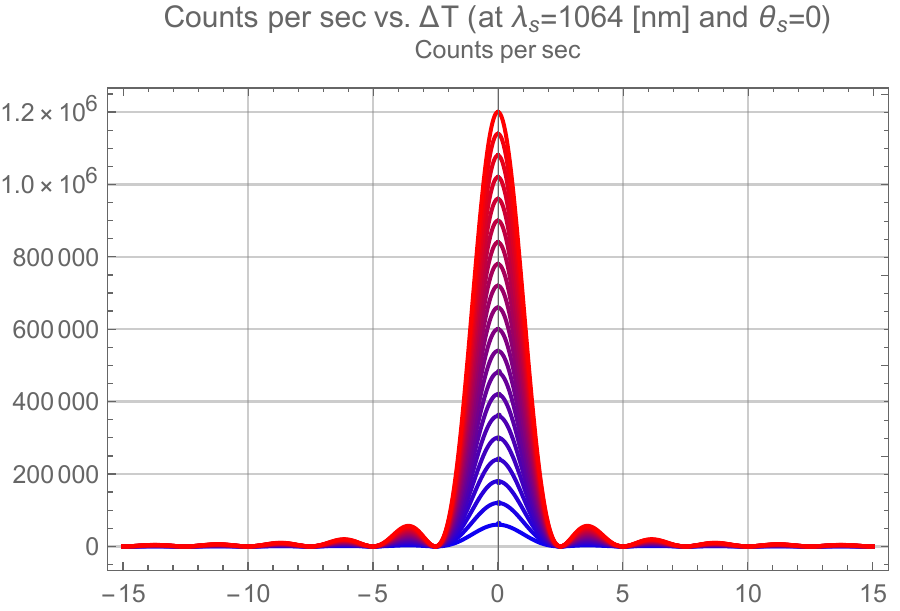}
    \includegraphics[width=0.33\textwidth]{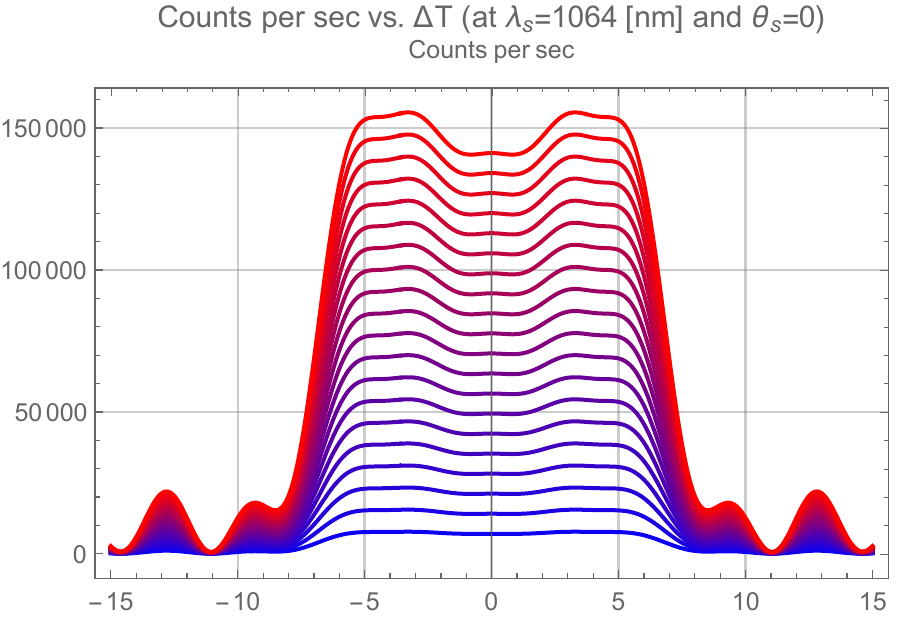}
    \includegraphics[width=0.33\textwidth]{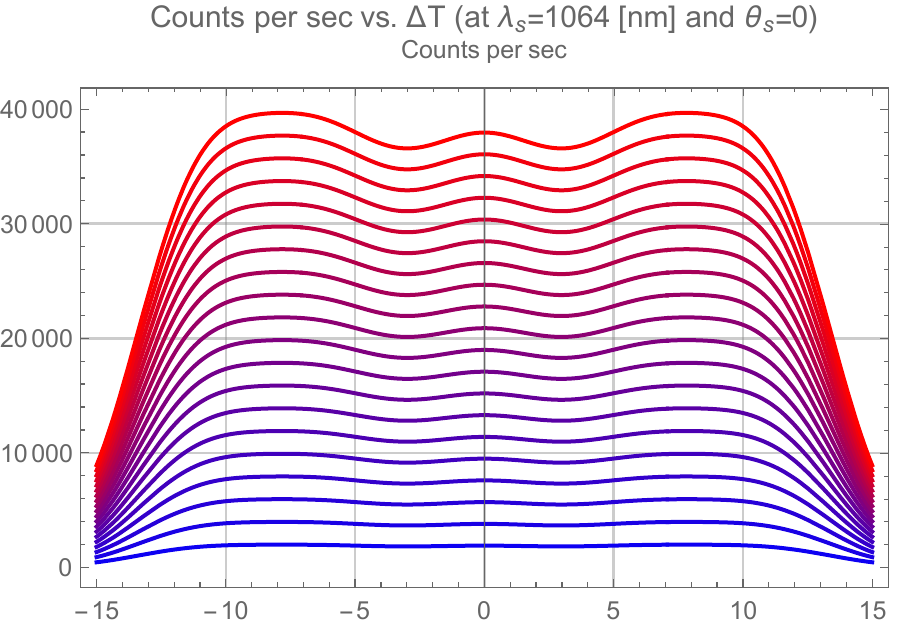}

    \begin{minipage}{1\linewidth}
    \raggedright (d) \hspace{156pt} (e) \hspace{156pt} (f)
    \end{minipage}

    \includegraphics[width=0.33\textwidth]{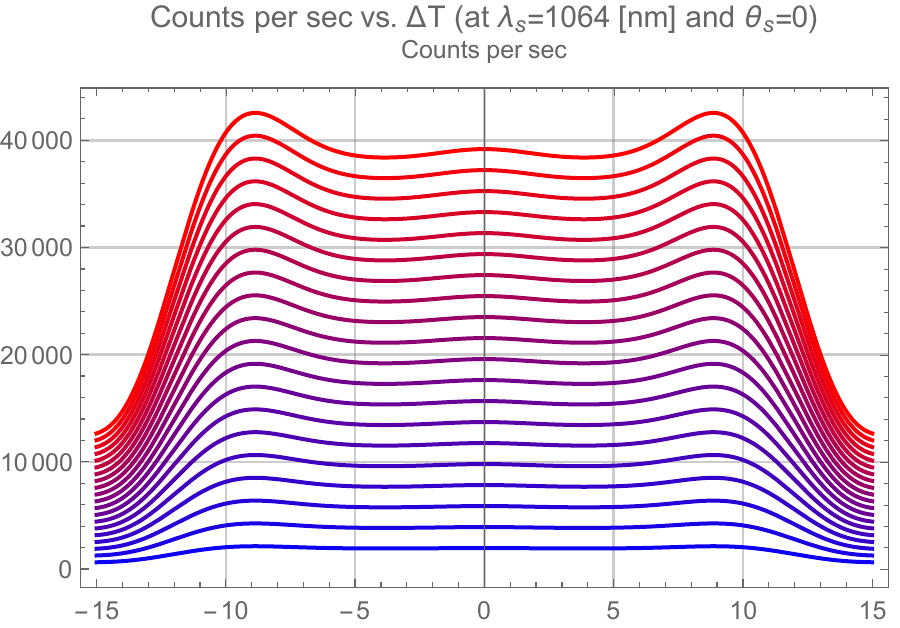}
    \includegraphics[width=0.33\textwidth]{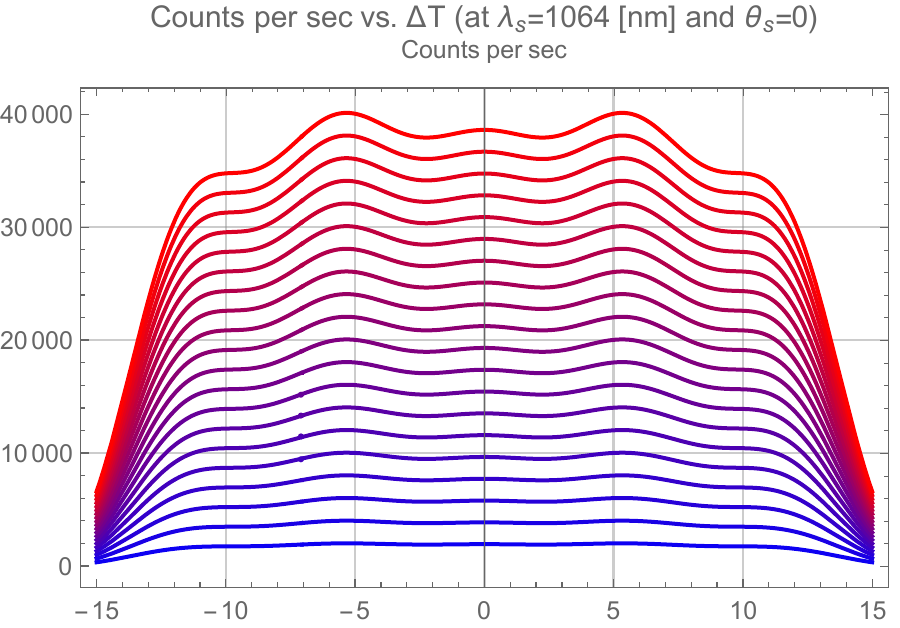}
    \includegraphics[width=0.33\textwidth]{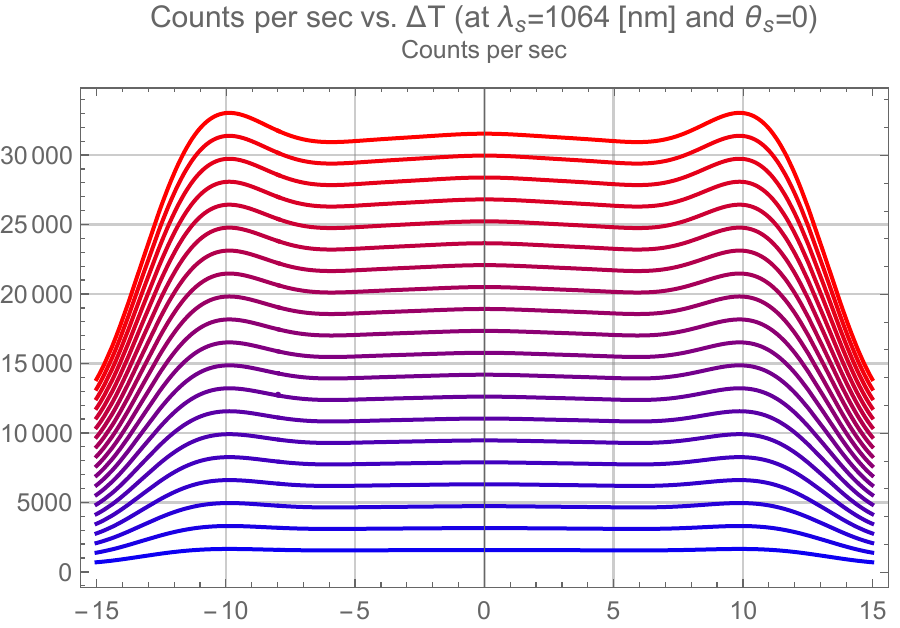}

    \begin{minipage}{1\linewidth}
    \raggedright (g) \hspace{156pt} (h) \hspace{156pt} (i)
    \end{minipage}

    \includegraphics[width=0.33\textwidth]{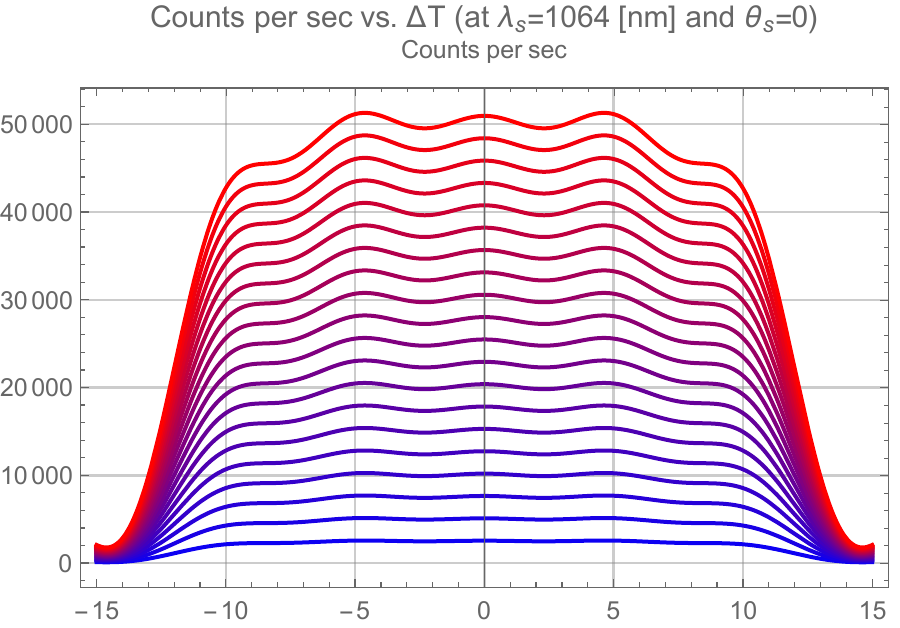}
    \includegraphics[width=0.33\textwidth]{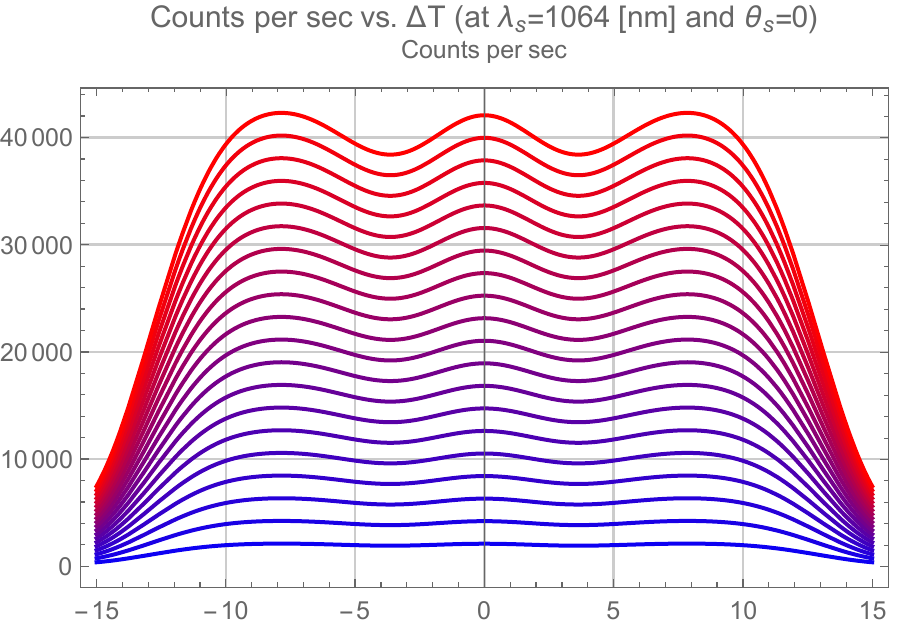}
    \includegraphics[width=0.33\textwidth]{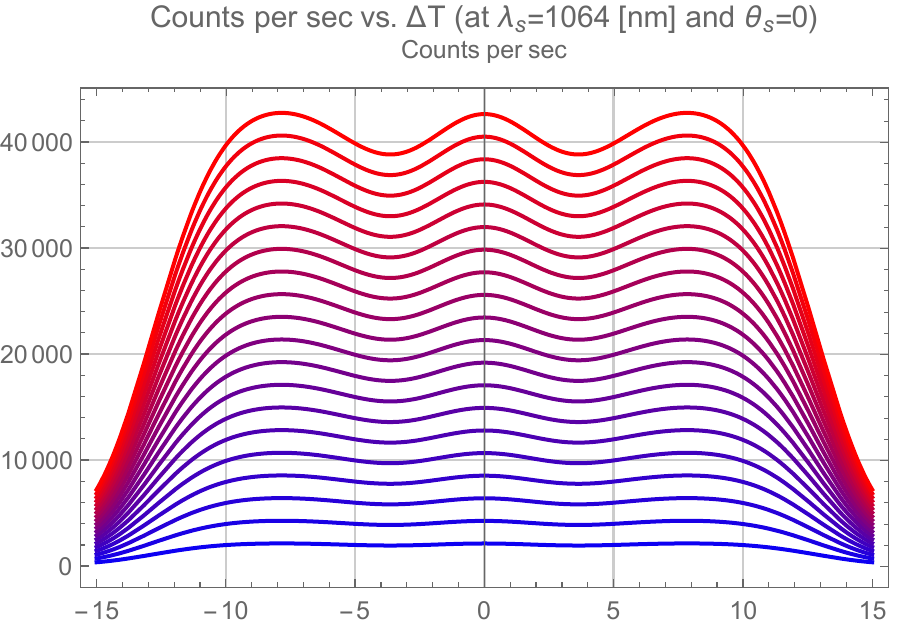}

     \caption{The number of generated entangled pairs per second as a function of the temperature deviation from $37^\circ C$ (at $\theta_s=0$) for 20 different incident powers of the pump of: (a) a perfectly phase matched crystal of length 20mm, (b,c,d,e,f,g,h,i) the DMCS crystals I,II,III,IV,V,VI,VII,VIII respectively. $\frac{\delta P_{\mathrm{pump}}}{P_{\mathrm{pump}}}$ from up to down in the graphs is $+1$ down to $-0.9$ in jumps of $-0.1$. The robustness of the process does not change under deviations in the incident pump's intensity.}
     \label{fig: S7 errors in kappa}
\end{figure*}

In the undepleted pump regime, the robustness we described before will not be violated due to the linearity of the dynamics. Errors of $\delta P_{\mathrm{pump}}$ in the pump's incident power cause $\delta P_{\mathrm{signal/idler}}$ as follows:
\begin{equation}
    \frac{\delta P_{\mathrm{pump}}}{P_{\mathrm{pump}}}= \frac{\delta P_{\mathrm{signal/idler}}}{P_{\mathrm{signal/idler}}}.
\end{equation}
We show in Fig. \ref{fig: S7 errors in kappa} the number of generated entangled pairs per second as a function of the temperature deviation from $37^\circ C$ (at $\theta_s=0$) for 20 different incident powers of the pump of: (a) a perfectly phase matched crystal of length 20mm, (b,c,d,e,f,g,h,i) the DMCS crystals I,II,III,IV,V,VI,VII,VIII respectively. $\frac{\delta P_{\mathrm{pump}}}{P_{\mathrm{pump}}}$ from up to down in the graphs is $+1$ down to $-0.9$ in jumps of $-0.1$. Unsurprisingly, the robustness of the process does not change under deviations in the incident pump's intensity.

\subsection{Different Pump Powers and Focusing Conditions}
\label{app: Different Pump Powers and Focusing Conditions}

\begin{figure*}[htbp]
    \centering
    \includegraphics[width=1\textwidth]{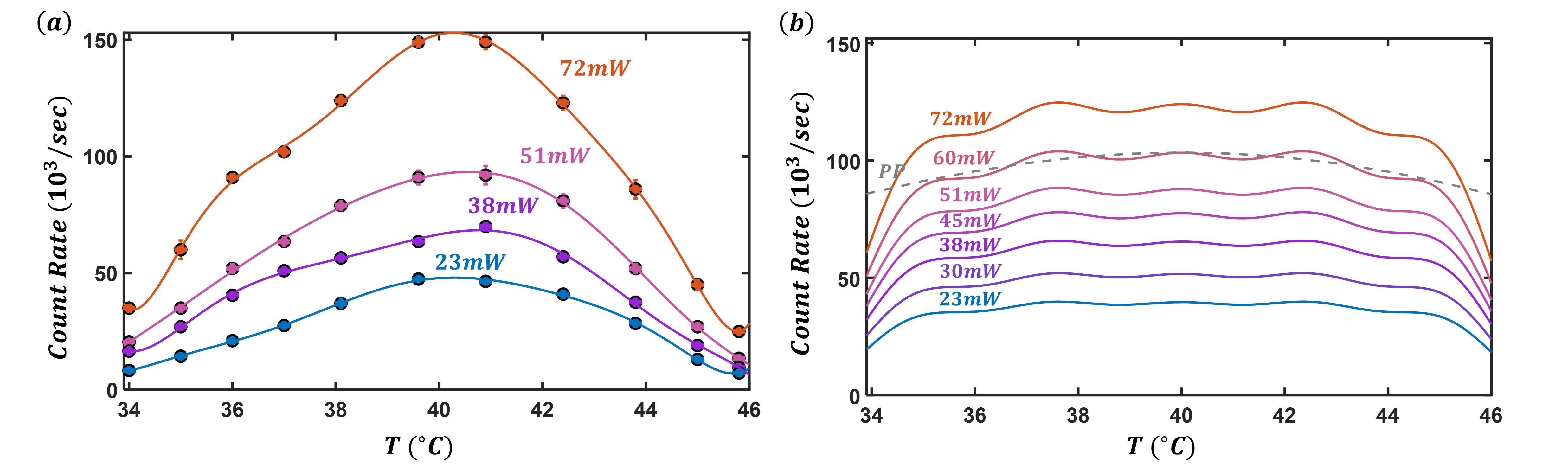}
    \caption{The photon pair generation rate of the DMCS design in different pump powers.
    $(a)$ Measured count rate vs. temperature for different pump powers varying in the range of $23mW-72mW$. 
    The pump power was decreased using a variable ND filter.
    $(b)$ Simulation count rate vs. temperature for different pump power varying from $23mW-72mW$. 
    In the dashed gray line, we see the simulated count rate for PP design with a length of $2mm$, ten times shorter than the equivalent composite design ($20mm$) but with pump power seventeen times stronger ($1.02W$ compared to $60mW$).}
    \label{fig: S8 nlinearity and BW equation}
\end{figure*}

Under the undepleted pump approximation, the output count rate scales linearly with the pump power, so robustness is preserved even when the input power is increased. 
Figures~\ref{fig: S8 nlinearity and BW equation}(a) and \ref{fig: S8 nlinearity and BW equation}(b) show the measured and simulated count rates at different pump powers. 
By normalizing the generated signals, we confirmed the linear dependence on pump power.

We also examined whether robustness comparable to DMCS can be achieved with a periodically poled (PP) crystal by shortening its length and increasing the pump power. 
Figure~\ref{fig: S8 nlinearity and BW equation}(b) illustrates that to obtain a robustness width similar to that of a $20\,\mathrm{mm}$ DMCS design pumped with $60\,\mathrm{mW}$, a PP crystal would need to be shortened to $2\,\mathrm{mm}$ and driven with $1.02\,\mathrm{W}$ pump power - seventeen times higher. 
Such high powers not only increase the risk of crystal damage but also dramatically enhance the generation of unwanted higher-order photon pairs. 

For completeness, we also considered the effect of optimal focusing. 
For a crystal of length $L$, the optimal beam waist is defined by $z_R = L/2$, yielding
\begin{equation}
    w_0 = \sqrt{\frac{L\lambda}{2\pi}}.
\end{equation}
With $\lambda = 532\,\mathrm{nm}$, the optimal waist values are
\begin{align}
    w_{0_\mathrm{DMCS}} &\approx 41 \,\mu\mathrm{m} \quad \text{for } L=20\,\mathrm{mm}, \\
    w_{0_\mathrm{PP}} &\approx 13 \,\mu\mathrm{m} \quad \text{for } L=2\,\mathrm{mm}.
\end{align}
This corresponds to a spot-size ratio of $\sqrt{10} \approx 3.16$, such that the short PP crystal can theoretically reach $\sim 10\times$ higher pump power density under optimal focusing. 
Even after accounting for this effect, the short PP crystal still requires approximately twice the pump intensity to match the efficiency of the DMCS design. 

We also note that achieving such tight focusing in practice introduces additional challenges, including increased alignment sensitivity, larger angular divergence ($\theta = \lambda/(\pi w_0)$), and reduced tolerance to incidence-angle drifts. 
For example, at optimal focusing our $20\,\mathrm{mm}$ DMCS device has $\theta \approx 4.1\,\mathrm{mrad}$, whereas the $2\,\mathrm{mm}$ crystal has $\theta \approx 13\,\mathrm{mrad}$ - a $3.2\times$ increase in angular spread. 
Thus, while the short PP crystal can in principle benefit from tighter focusing, in practice this configuration is more sensitive to alignment errors and less robust. 

In contrast, the DMCS design achieves high efficiency and robustness without requiring extreme focusing, which improves angular uniformity and increases tolerance to alignment and incidence-angle drifts. 
These results reinforce that the DMCS scheme provides a practical and efficient solution for generating entangled photons while minimizing unwanted higher-order pair generation, even when compared to optimally focused short PP crystals.

\subsection{Theoretical Joint Spectral Intensity Comparison}
\label{app: Theoretical Joint Spectral Intensity Comparison}

\begin{figure*}[htbp]
    \centering
    \includegraphics[width=1\textwidth]{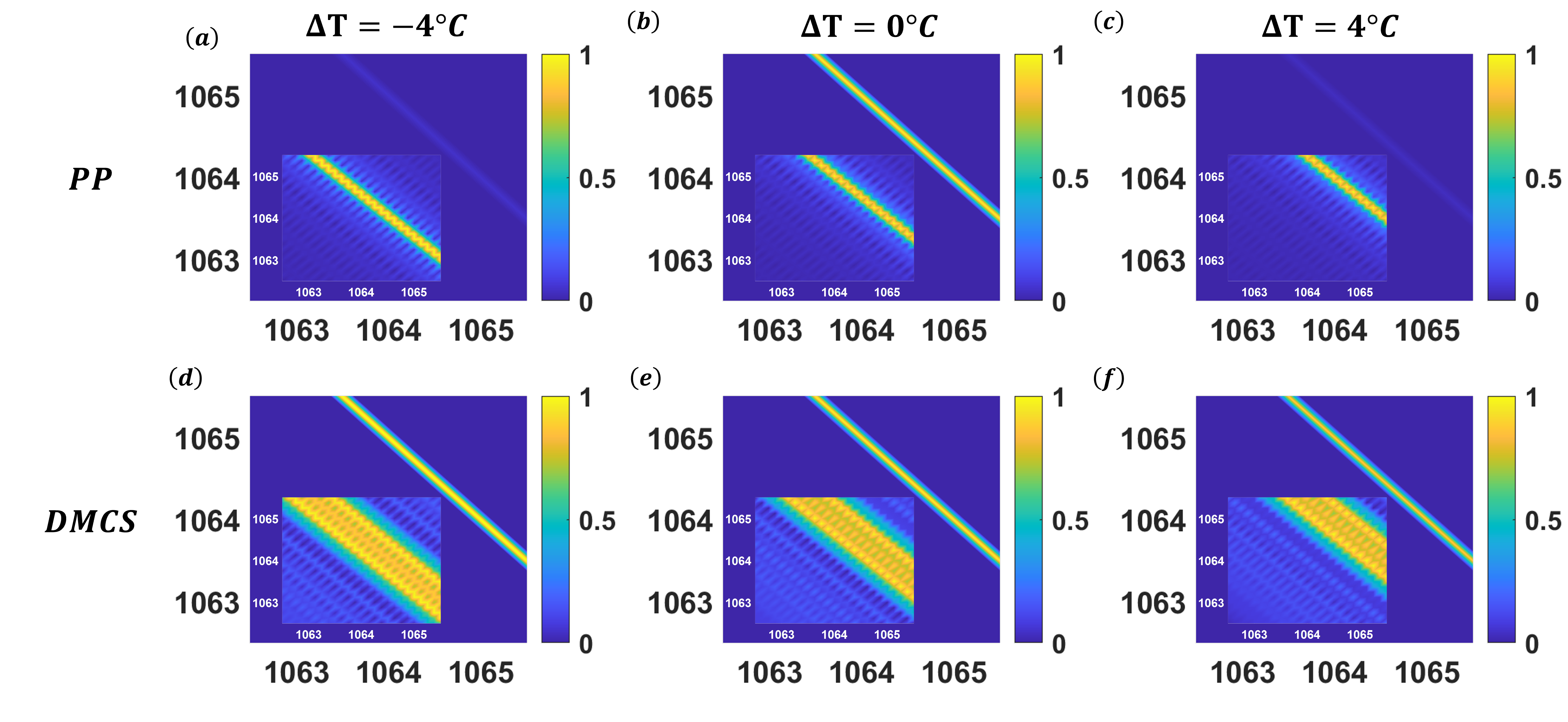}
    \caption{$(a)-(c)$ JSI of the periodically poled design at deviations of $-4^\circ C, 0^\circ C,+4^\circ C$) from the designated temperature inside the transfer window of the system's filter.
    $(d)-(f)$ JSI of the DMCS design at deviations of $-4^\circ C, 0^\circ C,+4^\circ C$) from the designated temperature.
    The insets are the phase-matching functions of each design and temperature deviation.}
    \label{fig: S9 JSI}
\end{figure*}

The joint spectral intensity can be calculated by taking the squared absolute value of the multiplication between the phase-matching function and the pump spectral amplitude. 
To demonstrate the theoretically superior robustness of the DMCS design over a periodically poled crystal, in Fig. \ref{fig: S9 JSI}(a)-(f) we present the JSI of both designs under different temperature deviations from the designated temperature. 
The JSIs of each scheme were normalized to the maximal value of the JSI at the designated temperature.
The insets are the absolute values of the phase-matching function of each design at each temperature.
The wider phase-matching function of the DMCS design allows it to remain stable over a wider temperature range.

\subsection{Efficiency-Robustness Trade-off Analysis}
\label{app: Efficiency-Robustness Trade-off Analysis}

In designing quantum photon sources, a fundamental consideration is the balance between the entangled photons generation efficiency and robustness. 
Our DMCS design intentionally prioritizes robustness to physical parameter variations while maintaining acceptable generation efficiency. 
This trade-off is often advantageous in quantum information applications where stability and predictability are more critical than raw efficiency, especially in applications where active stabilization methods (such as temperature control or feedback loops) would increase power consumption, mechanical complexity, and potential points of system failure. 
Our approach achieves robustness through passive structural elements, thereby offering a complementary strategy for enhanced system reliability.

Furthermore, since the SPDC process is linear in pump power under the undepleted pump approximation, as confirmed experimentally in Fig.~\ref{fig: S8 nlinearity and BW equation}, higher generation rates can be readily achieved by increasing the pump power while maintaining the enhanced stability of our design. 
This scaling provides a better trade-off compared to short PP designs and allows flexible performance optimization based on application requirements.

\section{Comparison With Chirped Design}
\label{app: Comparison With Chirped Design}

To provide a comprehensive comparison with alternative robust SPDC approaches, we designed and simulated a linearly chirped crystal optimized for broad spectral acceptance. 
This section presents the design parameters and performance characteristics of the chirped designs used in the quantitative comparison with our DMCS approach.
We selected two linearly chirped designs - \textbf{design I}, which is a naive linearly chirped design where at the edges the poling is designed to phase match the minimal and maximal wavelengths with effective conversion of the DMCS design, and \textbf{design II}, which is an optimized linearly chirped design.

\subsection{Design Methodology}
\label{app: Design Methodology}

The chirped pattern was designed using the following principles and equations. The poling period varies continuously along the crystal length according to the following equations:

\begin{equation}
K_g(z) = K_g(0) + \Delta K_g \cdot \frac{z}{L_c}
\end{equation}

\begin{equation}
\Lambda_{local}(z) = \frac{2 \pi}{K_g(z)}
\end{equation}
where $z$ is the position along the crystal, $K_g(0) = K_g(min)$, $\Delta K_g = K_g (max)-K_g (min)$, and $L_c$ is the total crystal length.
The polling direction is then calculated according  Eq. (\ref{eq: Poling for segmented}):
\begin{equation}
\chi^{(2)}(z) = |\chi^{(2)}| \cdot sign \left[ cos \left( K_g(0) z + \Delta K_g \cdot \frac{z^2}{2L_c} \right) \right]
\end{equation}

\subsection{Optimization for the Problem with Linear Mismatch/Chirp design}
\label{app: Optimization for the Problem with Linear Mismatch/Chirp design}

We consider the situation where the phase mismatch varies linearly with the propagation coordinate.  
The mismatch is written as
\[
\Delta k = A + Bz ,
\]
where the constants $A$ and $B$ characterize the offset and slope of the variation.
To implement such a design under the manufacturing limitations described in Sec.\ref{app: Polings for The Designs and Manufacturing Limitations}, we employ Eq.(\ref{eq: Poling for segmented}), where $\Delta k_{\text{pol}}$ represents the linear phase mismatch $\Delta k$ shifted by the crystal wave number corresponding to the phase-matched process.
For convenience we introduce $a = A$ and $b = \tfrac{1}{2}B$.  

The coupled–mode dynamics can then be expressed in matrix form as
\begin{equation}
\frac{\partial}{\partial z} 
\begin{bmatrix}
A_s(z) \\
A_i^*(z)
\end{bmatrix}
=
\begin{bmatrix}
0 & -i \kappa e^{-i(a+bz)z} \\
i \kappa e^{i(a+bz)z} & 0
\end{bmatrix}
\begin{bmatrix}
A_s(z) \\
A_i^*(z)
\end{bmatrix}.
\end{equation}

The exact solution has the form
\begin{equation}
\begin{bmatrix}
A_s(z) \\
A_i^*(z)
\end{bmatrix}
=
\begin{bmatrix}
\alpha(z) & \beta(z) \\
\beta^*(z) & \alpha^*(z)
\end{bmatrix}
\begin{bmatrix}
A_s(0) \\
A_i^*(0)
\end{bmatrix},
\end{equation}
where the coefficients $\alpha(z)$ and $\beta(z)$ are expressed through special functions.

\begin{widetext}
\scalebox{0.76}{
\parbox{\linewidth}{
\begin{subequations}
    \begin{gather}
\begin{split}
& \alpha(z) = e^{-iz(a+bz)} \\
&\frac{
 2b  {}_1F_1 \left(\tfrac{1}{2}-\tfrac{i\kappa^2}{4b}, \tfrac{1}{2}, \tfrac{i(a+2bz)^2}{4b}\right)
 \Big[ 2b  {}_1F_1 \left(1-\tfrac{i\kappa^2}{4b}, \tfrac{1}{2}, \tfrac{ia^2}{4b}\right) 
 - i a^2  {}_1F_1 \left(1-\tfrac{i\kappa^2}{4b}, \tfrac{3}{2}, \tfrac{ia^2}{4b}\right)\Big]
 - a(a+2bz)\kappa^2  {}_1F_1 \left(\tfrac{1}{2}-\tfrac{i\kappa^2}{4b}, \tfrac{3}{2}, \tfrac{ia^2}{4b}\right) 
   {}_1F_1 \left(1-\tfrac{i\kappa^2}{4b}, \tfrac{3}{2}, \tfrac{i(a+2bz)^2}{4b}\right)
}{
4b^2  {}_1F_1 \left(\tfrac{1}{2}-\tfrac{i\kappa^2}{4b}, \tfrac{1}{2}, \tfrac{ia^2}{4b}\right) 
  {}_1F_1 \left(1-\tfrac{i\kappa^2}{4b}, \tfrac{1}{2}, \tfrac{ia^2}{4b}\right)
- a^2\Big(2ib  {}_1F_1 \left(\tfrac{1}{2}-\tfrac{i\kappa^2}{4b}, \tfrac{1}{2}, \tfrac{ia^2}{4b}\right) 
+ \kappa^2  {}_1F_1 \left(\tfrac{1}{2}-\tfrac{i\kappa^2}{4b}, \tfrac{3}{2}, \tfrac{ia^2}{4b}\right)\Big) 
  {}_1F_1 \left(1-\tfrac{i\kappa^2}{4b}, \tfrac{3}{2}, \tfrac{ia^2}{4b}\right)
}.
\end{split}
\\
\beta(z) = \frac{2b   e^{-iz(a+bz)} \kappa \Big[ -H_{-1+\frac{i\kappa^2}{2b}} \left(\frac{(-1)^{1/4}(a+2bz)}{2\sqrt{b}}\right) {}_1F_1 \left(\tfrac{1}{2}-\tfrac{i\kappa^2}{4b}, \tfrac{1}{2}, \tfrac{ia^2}{4b}\right) \notag + H_{-1+\frac{i\kappa^2}{2b}} \left(\frac{(-1)^{1/4}a}{2\sqrt{b}}\right) {}_1F_1 \left(\tfrac{1}{2}-\tfrac{i\kappa^2}{4b}, \tfrac{1}{2}, \tfrac{i(a+2bz)^2}{4b}\right) \Big]} {2(-1)^{1/4}\sqrt{b}(2ib+\kappa^2) H_{-2+\frac{i\kappa^2}{2b}} \left(\frac{(-1)^{1/4}a}{2\sqrt{b}}\right) {}_1F_1 \left(\tfrac{1}{2}-\tfrac{i\kappa^2}{4b}, \tfrac{1}{2}, \tfrac{ia^2}{4b}\right) + a H_{-1+\frac{i\kappa^2}{2b}} \left(\frac{(-1)^{1/4}a}{2\sqrt{b}}\right) \big(-2b {}_1F_1 \left(\tfrac{1}{2}-\tfrac{i\kappa^2}{4b}, \tfrac{1}{2}, \tfrac{ia^2}{4b}\right) + i\kappa^2 {}_1F_1 \left(\tfrac{1}{2}-\tfrac{i\kappa^2}{4b}, \tfrac{3}{2}, \tfrac{ia^2}{4b}\right)\big)}.
    \end{gather}
\end{subequations}
}
}
\end{widetext}
\normalsize
Here, ${}_1F_1$ denotes the confluent hypergeometric function and $H_\nu$ is the Hermite function of order $\nu$.  

When the parameter $\abs{\kappa^2 / 2b}$ is very small, which is the case in essentially all practical chirped designs, one may simplify the solution by expanding to first order in $\kappa/\sqrt{2b}$. In this approximation the coefficients reduce to
\begin{widetext}
\begin{subequations}
\begin{align}
\alpha(z) &= 1 + O  \left(\frac{\kappa^2}{2b}\right), \\
\beta(z) &= e^{i\left(\frac{a^2}{4b}+\pi/4\right)} 
\sqrt{\frac{\pi}{2}}  \frac{\kappa}{\sqrt{2b}}
\Bigg[ \erf  \left(\frac{e^{i\pi/4} a}{2\sqrt{b}}\right)
- \erf  \left(\frac{e^{i3\pi/4}(a+2bz)}{2\sqrt{b}}\right) + O  \left(\frac{\kappa^2}{2b}\right) \Bigg] .
\end{align}
\end{subequations}
\end{widetext}
In fact, the optimization process can be carried out using the exact solution without assuming $\abs{\kappa^2/2b} \ll 1$. However, adopting this assumption and simplifying the expression for $\beta(z)$ makes simulations significantly easier and faster.

In practice, variations in temperature, wavelength, or incident angle are effectively captured by changes in the parameter $a$. 
The performance of the structure can therefore be quantified by evaluating
\[
\mu = \left|\beta(L_{\text{tot}}=0.02)\right|^2 ,
\]
as a function of $a$ for different choices of $b$.  

The optimization goal is twofold. First, the value of $\mu$ should remain nearly constant over a wide range of $a$, and in particular it should not fall below ninety percent of its maximum. Second, the maximal $\mu$ should exceed fifteen percent of the value obtained in a periodically poled crystal, for which
\[
\mu_{\text{PP}} = \sinh^2(\kappa L_{\text{tot}}).
\]

To achieve this, the parameter $b$ is scanned starting from $10^2$, which is set by the requirement $\abs{\kappa^2/2b} \ll 1$, up to $10^8$, which is imposed by the minimal feasible poling length of approximately three micrometers.
We named the most robust chirped scheme satisfying the above criteria design II.

\subsection{Chirped Design Parameters}
\label{app: Chirped Design Parameters}

The chirped designs were using the following parameters:

\begin{table}[htbp]
\centering
\begin{tabular}{|l|c|c|}
\hline
\textbf{Parameter} & \textbf{Value design I} & \textbf{Value design II} \\
\hline
$K_g$ (min) & 700,609 [$m^{-1}$] & 701,819 [$m^{-1}$] \\
\hline
$K_g$ (max) & 702,546 [$m^{-1}$] & 703,298 [$m^{-1}$] \\
\hline
Crystal length & 20 mm & 20 mm \\
\hline
Operating temperature & 37 °C & 37 °C \\
\hline
Pump wavelength & 532 nm & 532 nm \\
\hline
\end{tabular}
\caption{Design parameters for the chirped crystals used in the comparative analysis.}
\label{tab:chirped_parameters}
\end{table}

\subsection{Results and Performance Characteristics}
\label{app: Results and Performance Characteristics}

\begin{figure*}[htbp]
    \centering
    \includegraphics[width=1\textwidth]{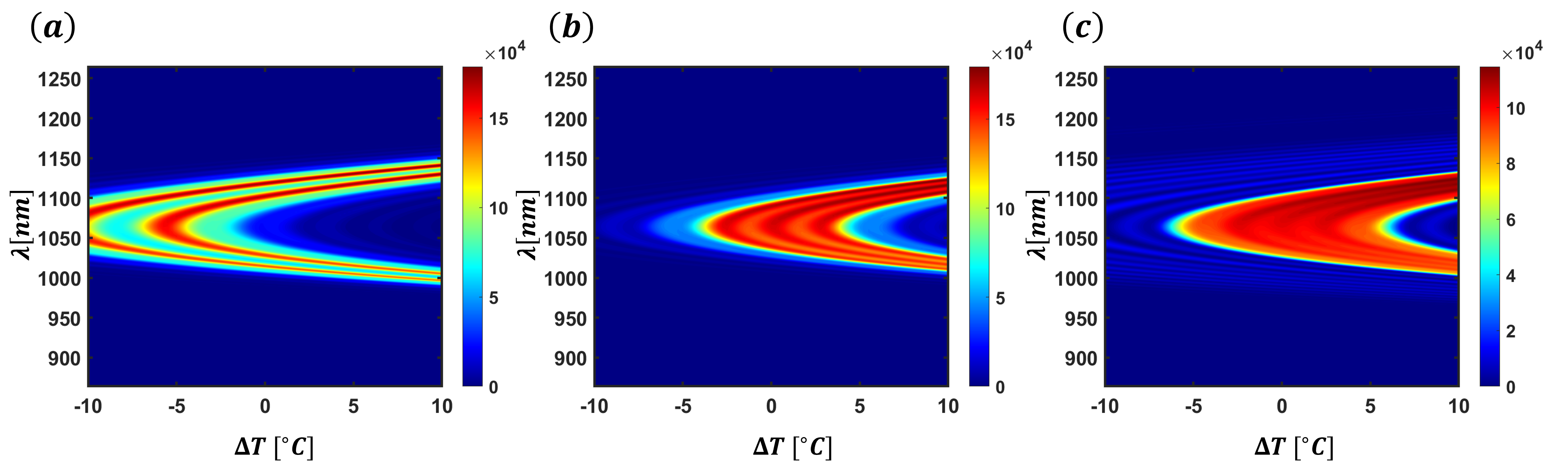}
    \caption{Performance comparison between linearly chirped designs and DMCS design: Count rate vs. temperature variation and signal wavelength for (a) chirped design I --- linear chirp for similar wavelength range to that of the DMCS design, (b) chirped design II --- linear chirp optimized for a wide robustness width in temperature, (c) DMCS design VI.}
    \label{fig: S10 chirped_comparison}
\end{figure*}

\begin{figure*}
    \centering

    \begin{minipage}{1\linewidth}
    \raggedright (a) \hspace{240pt} (b)
    \end{minipage}
    
    \includegraphics[width=0.49\linewidth]{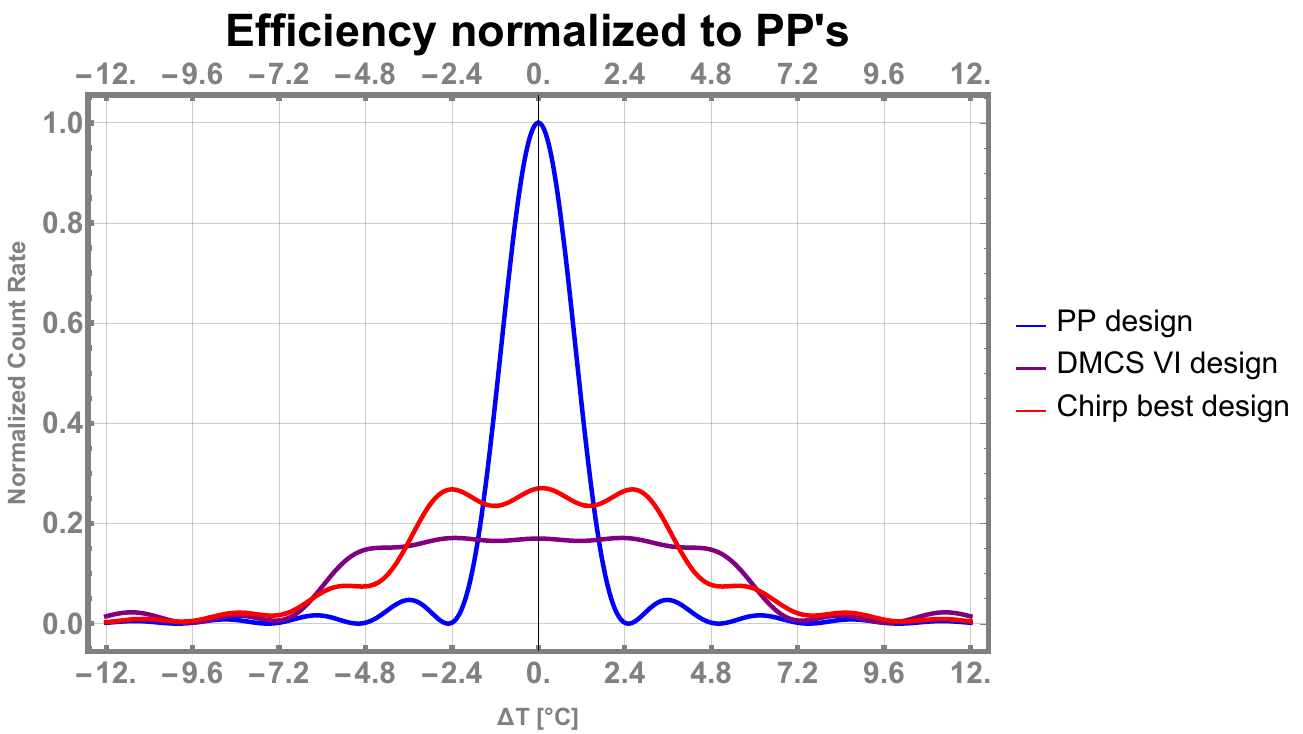}
    \includegraphics[width=0.49\linewidth]{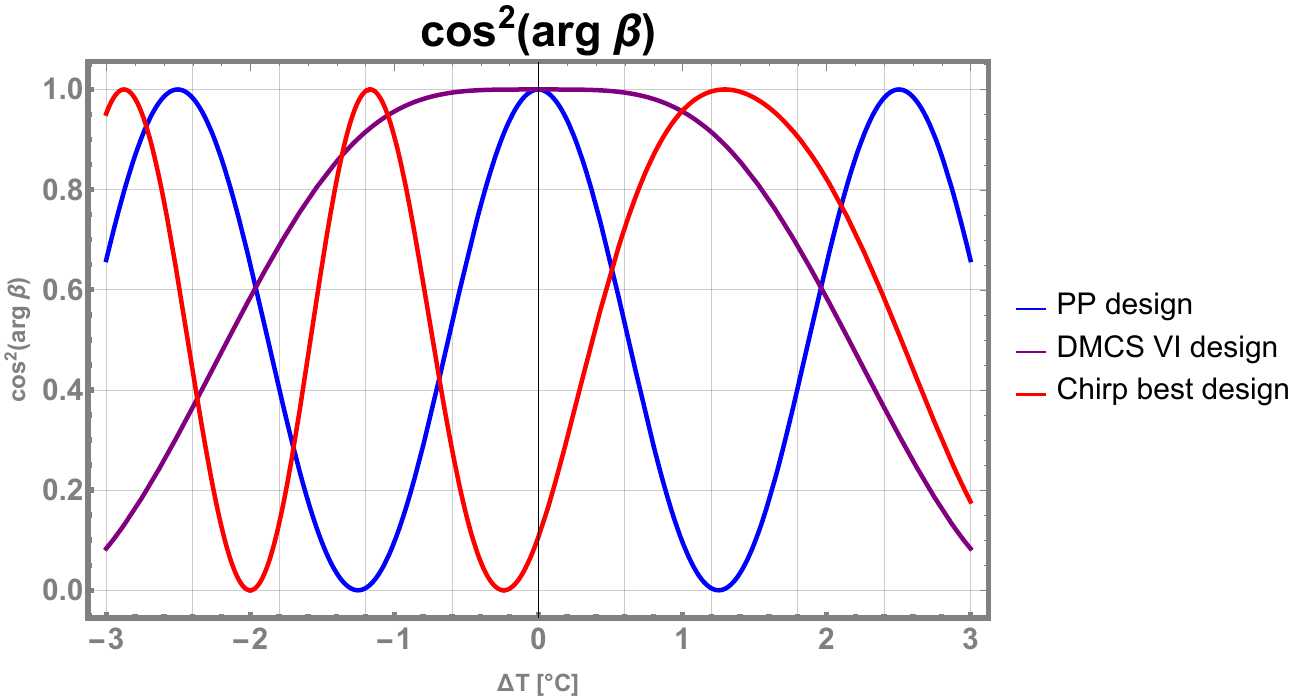}
    
    \caption{Performance comparison for signal at 1064 nm between a PP design, the best chirped design (chirped design II), and our DMCS design VI:
    (a) Count rate vs. temperature variation, (b) Phase vs. temperature variation.}
    \label{fig: S11 1DcomparisonWithChirp}
\end{figure*}

Figure~\ref{fig: S10 chirped_comparison} shows the temperature and wavelength dependent performance of the chirped designs compared to our DMCS approach. 
Figure~\ref{fig: S11 1DcomparisonWithChirp}(a) compares the performance of the best chirped design (design II) with our DMCS design VI and with a PP design.
Figure~\ref{fig: S11 1DcomparisonWithChirp}(b) demonstrates phase robustness to variations in temperature by ploting $cos^2 \left( arg \left( \beta \right) \right)$ temperature variations for design II, DMCS design VI, and a PP design.

The chirped designs demonstrated the following characteristics:

\begin{itemize}
    \item \textbf{Spectral bandwidth:} Efficient conversion across the wavelength range of 1012 nm-1123 nm (design I) and 1023 nm-1110 nm (design II).
    \item \textbf{Temperature acceptance:} Operating range of -15 °C to -2 °C for design I, and approximately $\pm$3.5 °C at 90\% efficiency for design II.
    \item \textbf{Count rate uniformity:} Variations of $\sim$60\% for design I and $\sim$10\% for design II across their operating range.
    \item \textbf{Peak count rate:} Approximately $1.6\cdot 10^5$ cps at 1064 nm for design I and $1.64\cdot 10^5$ cps for design II.
    \item \textbf{Spectral phase:} Complex, since each wavelength is generated in different regions of the crystal and experiences different dispersion (see Ref. [19] in the main text).
    \item \textbf{Temperature dependent phase:} Displays nontrivial and unstable behavior.
\end{itemize}

\subsection{Comparative Analysis with DMCS}
\label{app: Comparative Analysis with DMCS}

Our DMCS approach offers several distinct advantages:

\begin{enumerate}
    \item \textbf{Uniformity and robustness:} DMCS exhibits superior stability in both count-rate and phase against temperature variations, with typical count-rate variations below 10\% across $\pm$5 °C, compared to chirped designs where variation is higher and robustness width narrower.
    \item \textbf{Temporal walk-off and phase issues:} Chirped designs generate photon pairs predominantly in limited regions of the crystal, leading to temporal mismatch and complex phase behavior. 
    In contrast, DMCS generates photon pairs throughout the propagation, enabling operation with shorter crystals, reduced walk-off, and a spectral phase that can be effectively compensated.
    \item \textbf{Crystal length flexibility:} DMCS can be implemented across a wide range of lengths (mm-cm), whereas chirped designs typically require long crystals to achieve efficient conversion.
    \item \textbf{Efficiency at low pump powers:} Chirped designs often suffer efficiency reduction due to distributed phase-matching, while DMCS maintains high efficiency across power regimes.
\end{enumerate}

Although chirped designs show slightly higher peak count-rates, this advantage can be compensated for by higher pump power. 
Their main strength lies in achieving very broad spectral acceptance, albeit at the cost of long-crystal requirements, efficiency non-uniformity, and unstable phase. 
In contrast, DMCS enables compact sources with limited pump power, uniform efficiency, and a stable phase that can be readily compensated, making it the more practical choice for many quantum optics applications.

\end{document}